\documentclass[manuscript,screen]{acmart}
\usepackage{multirow}
\usepackage{makecell}
\usepackage{subcaption}
\usepackage[ruled,linesnumbered]{algorithm2e}
\usepackage{graphicx}
\usepackage{caption}
\usepackage{threeparttable}
\usepackage{xcolor}
\usepackage{array}
\newcolumntype{Z}{>{\centering\arraybackslash}m{0.115\linewidth}}

\AtBeginDocument{%
  }
\setcopyright{acmlicensed}
\copyrightyear{2018}
\acmYear{2018}
\acmDOI{XXXXXXX.XXXXXXX}
  
\acmISBN{978-1-4503-XXXX-X/2018/06}

\begin{document}

\title{Differentiable Routability-Driven Package Floorplanning with Pin Assignment}

\author{Yiqi Huang}
\email{huang170825@126.com}
\affiliation{%
  \institution{Fuzhou University}
  \city{Fuzhou}
  \state{Fujian}
  \country{China}
  \postcode{350116}
}

\author{Zepeng Li}
\email{li\_zepeng@yeah.net}
\affiliation{%
  \institution{Fuzhou University}
  \city{Fuzhou}
  \state{Fujian}
  \country{China}
  \postcode{350116}
}

\author{Zhen Zhuang}
\email{zhuangzhen1995@gmail.com}
\affiliation{%
  \institution{The Chinese University of Hong Kong}
  \country{Hong Kong SAR}
  \postcode{999077}
}

\author{Kehao Chen}
\email{zeroeeau@gmail.com}
\affiliation{%
  \institution{Fuzhou University}
  \city{Fuzhou}
  \state{Fujian}
  \country{China}
  \postcode{350116}
}

\author{Genggeng Liu}
\authornote{Corresponding author: Genggeng Liu.}
\email{liugenggeng@fzu.edu.cn}
\affiliation{%
  \institution{Fuzhou University}
  \city{Fuzhou}
  \state{Fujian}
  \country{China}
  \postcode{350116}
}

\author{Tsung-Yi Ho}
\email{tyho@cse.cuhk.edu.hk}
\affiliation{%
  \institution{The Chinese University of Hong Kong}
  \country{Hong Kong SAR}
  \postcode{999077}
}

\renewcommand{\shortauthors}{Y. Q. Huang, et al.}

\begin{abstract}
As advanced packaging technology evolves, the increasing interconnect density of signal nets in redistribution layers (RDLs) makes routability critical for package floorplanning. On the other hand, power integrity requirements often reserve fan-in regions for the power delivery network (PDN), leaving signal nets to be routed through fan-out regions. This makes fan-in regions unavailable to signal nets and makes routability estimation during package floorplanning particularly challenging. Existing uniform grid-based congestion models cannot accurately characterize fan-out region congestion, and previous pin assignment methods have difficulty in evaluating net crossings.
To address these issues, we propose a differentiable routability-driven floorplanning with pin assignment algorithm for advanced packaging under the fan-out routing style.
First, we propose a differentiable wirelength minimization method that directly models discrete chip orientations and back-propagates wirelength gradients to chip locations and orientations, reducing wirelength under fixed pin selection while avoiding the bias of continuous-angle modeling.
In addition, we propose a crossing-aware pin assignment method that incorporates net-crossing cost into a multi-strategy DPSO optimization algorithm and adopts GPU-parallel cost evaluation, effectively reducing wirelength within a short runtime.
Finally, we propose a differentiable routability maximization method that constructs a congestion estimation model tailored to the fan-out routing style and establishes a back-propagation path from congestion information to chip locations, thereby guiding routability optimization.
Experimental results show that the proposed method achieves 100\% routability on all benchmarks. For cases successfully routed by the baselines, it reduces wirelength by a maximum of approximately 23\% compared with a leading floorplanning method equipped with the proposed pin assignment flow.
\end{abstract}



\received{20 February 2007}
\received[revised]{12 March 2009}
\received[accepted]{5 June 2009}
\maketitle

\section{Introduction}
As advanced technology node approaches physical limits, sustaining Moore's Law has become increasingly difficult~\cite{ramm20103d,heinig2014system}. The economic benefit of adding more transistors to a single chip has diminished, leaving few wafer fabs able to absorb the associated manufacturing costs~\cite{zhuang2022multi}. To meet demands for high performance while reducing manufacturing cost, manufacturers such as TSMC, Samsung, and Intel are investing heavily in advanced packaging~\cite{zhuang2022multi}. As one of the representative advanced packaging technologies, fan-out wafer-level packaging (FOWLP) has shown great promise~\cite{yu2015new,ouyang2023fan,zoschke2024key}. FOWLP improves I/O density, energy efficiency~\cite{liang2025multi,guan2017fowlp,wang2019power,boon2024multi}, and thermal performance~\cite{cardoso2015thermally,kim2020effective}, making it a key technology for high-performance computing and artificial intelligence systems.
\begin{figure*}[t!]
	\centering
	\begin{subfigure}[b]{0.73\textwidth}
		\centering
		\includegraphics[width=\linewidth]{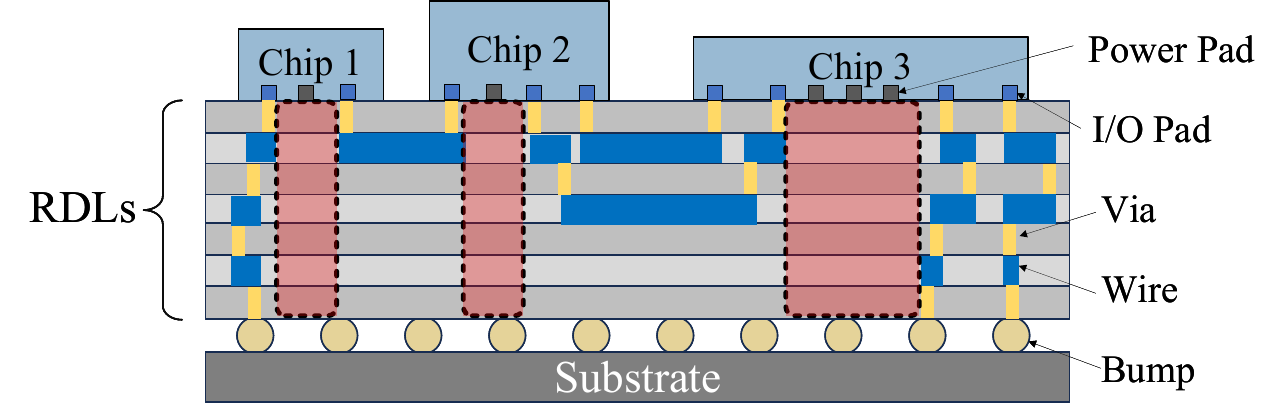}
		\caption{}
		\label{f:1a}
	\end{subfigure}
	\hspace{0.005\linewidth}	
	\begin{subfigure}[b]{0.25\textwidth}
		\centering
		\includegraphics[width=\linewidth]{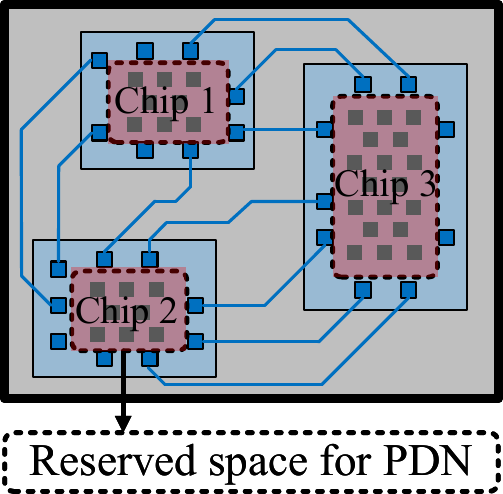}
		\caption{}
		\label{f:1b}
	\end{subfigure}
	\hfill

	\caption{Typical structure of an FOWLP. (a) Cross-sectional view. (b) Top view.}
	\label{f:1}
\end{figure*}

Figure~\ref{f:1}(\subref{f:1a}) shows a typical cross section of an FOWLP. Multiple heterogeneous chips are placed on the top side of the package. Beneath these chips, multiple redistribution layers are deployed to redistribute signal connections. These RDLs consist of stacked via layers and metal layers. Wires are routed on the metal layers, while vias are utilized to connect wires across different metal layers. At the bottom of the package, the RDLs are further connected to the package substrate through bump. In RDLs, minimizing chip-to-chip connection length is critical because it directly affects package-level performance. Package floorplanning must therefore place chips in routability relative positions, allowing signal nets to use shorter redistribution-layer routes. In addition, the distribution of routing resources across different regions of the RDL must be considered to ensure the routability of the final floorplanning result.
Figure~\ref{f:1}(\subref{f:1b}) illustrates the fan-out routing style considered in this work. The region under a chip projection is treated as the fan-in region, whereas the region outside the chip projection is treated as the fan-out region. I/O pads are distributed along the fan-in boundaries, with signal nets escaping from the chip toward the
fan-out region for chip-to-chip connections. To ensure reliable power delivery, the power delivery network (PDN) often needs to traverse the limited fan-in region, where power pads are located, using wider wires to support high current density and reduce IR drop~\cite{lu2024power,yip2022reliability}. Therefore, Li et al.~\cite{li2026redistribution} guide signal nets through fan-out regions to reduce fan-in usage. This makes it necessary for package floorplanning to explicitly distinguish fan-in and fan-out regions when estimating routability.
\begin{figure*}[t!]
	\centering
		\includegraphics[width=1\linewidth]{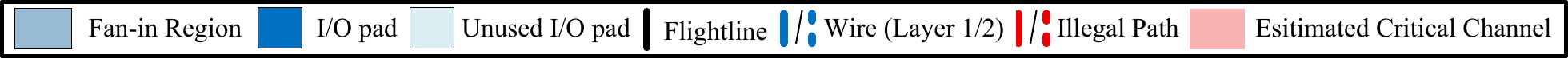}\par
		\vspace{1mm}
	\begin{subfigure}[b]{0.33\textwidth}
		\centering
		\includegraphics[width=\linewidth]{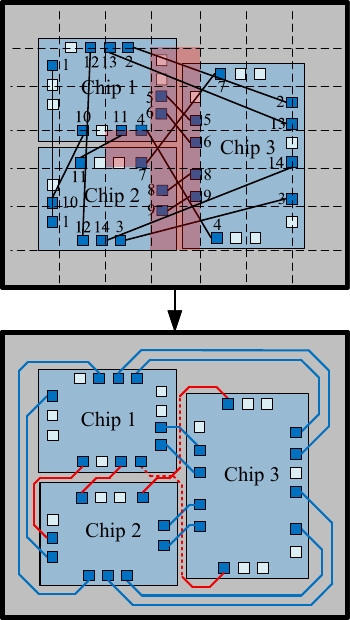}
		\caption{}
		\label{f:2a}
	\end{subfigure}
	\hfill
	\begin{subfigure}[b]{0.33\textwidth}
		\centering
		\includegraphics[width=\linewidth]{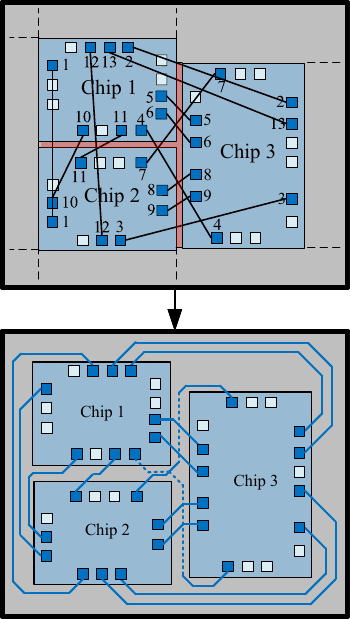}
		\caption{}
		\label{f:2b}
	\end{subfigure}
	\hfill
	\begin{subfigure}[b]{0.33\textwidth}
	\centering
	\includegraphics[width=\linewidth]{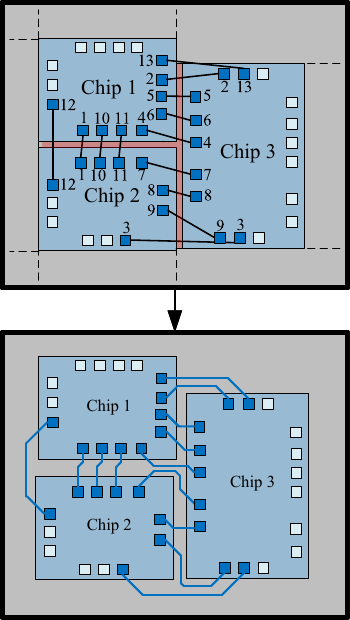}
	\caption{}
	\label{f:2c}
	\end{subfigure}
	\caption{Comparison of routability optimization flows. (a) The existing flow uses uniform-grid-based congestion estimation, which may treat unavailable fan-in regions as routing resources and lead to underestimated congestion and routing failure. (b) The proposed fan-out-aware congestion estimation model explicitly distinguishes fan-in and fan-out regions, enabling critical channels to be enlarged for better routability.
	(c) Based on (b), pin assignment is further introduced to assign the pins of each net to closer locations, thereby further reducing wirelength while maintaining routability.
	}
	\label{f:2}
\end{figure*}

Existing routability-driven floorplanning flows usually consist of two stages. They first perform wirelength-driven floorplanning to obtain a legal solution with short wirelength, and then apply routability optimization to enlarge critical routing channels so that the final layout can be successfully routed. Figure~\ref{f:2} illustrates the process from a wirelength-optimized solution to the routed result after routability optimization. In previous flows, chip translation and rotation are optimized to obtain a short-wirelength legal solution. Based on this solution, congestion is typically estimated using a uniform-grid-based model, and routability optimization is then performed according to the estimated congestion. However, such methods do not explicitly distinguish fan-in and fan-out regions, and may treat fan-in regions as available routing resources. Therefore, when the RDL router tends to guide signal nets through fan-out regions, these methods may produce unreliable congestion estimates. As shown in Figure~\ref{f:2}(\subref{f:2a}), the available routing resources estimated by the uniform-grid-based model include a large portion of unavailable fan-in regions. As a result, the subsequent routability optimization may incorrectly assume that sufficient routing resources exist in this region and fail to effectively enlarge the critical routing channel, leaving no feasible routing path for some nets on that layer. In contrast, as shown in Figure~\ref{f:2}(\subref{f:2b}), the proposed fan-out-aware congestion estimation model explicitly distinguishes fan-in and fan-out regions, allowing the optimizer to accurately identify congested fan-out channels and enlarge critical routing channels for better routability. On this basis, we further introduce pin assignment to reduce wirelength. As shown in Figure~\ref{f:2}(\subref{f:2c}), by assigning nets to more appropriate pins, the flow incorporating pin assignment can obtain a solution with shorter wirelength.

The major contributions of this work are summarized as follows.

(1) To the best of our knowledge, this paper is the first to formulate the routability-driven package floorplanning problem under the fan-out routing style. To address this problem, we propose a GPU-friendly three-stage framework that aims to improve routability while reducing wirelength.

(2) We propose a differentiable wirelength minimization method that directly models legal discrete chip orientations and back-propagates wirelength gradients to chip locations and orientations, thereby avoiding the dependence on the initialization of continuous rotation angles in indirect orientation modeling and reducing wirelength under fixed pin selection.

(3) We propose a crossing-aware pin assignment method that incorporates potential net-crossing cost into a multi-strategy DPSO algorithm and uses GPU-parallel cost evaluation, efficiently generating pin assignment solutions with lower HPWL within a short runtime.

(4) We propose a differentiable routability maximization method based on a fan-out-aware congestion estimation model. By partitioning fan-out routing regions and estimating local routing demand with a Top-K path algorithm, this model characterizes congestion in fan-out regions. The estimated congestion is then formulated as a differentiable congestion loss. The gradients of this loss are back-propagated through the routing-channel lengths to the chip locations, guiding the optimizer to enlarge critical routing channels and improve RDL routability.

(5) Experimental results show that the proposed method achieves 100\% routability on all benchmarks. Moreover, on fully routable cases, it reduces routed wirelength by a maximum of approximately 23\% compared with a leading floorplanning method equipped with the proposed pin assignment flow.

The remainder of this article is organized as follows. Section 2 reviews related work. Section 3 presents the problem formulation and related definitions. Section 4 describes the design and implementation of the proposed algorithm. Section 5 reports the experimental results. Section 6 concludes this article.

\section{Related Work}

The studies most relevant to this work can be divided into two categories: package floorplanning and pin assignment. Although pin assignment methods are not always developed specifically for package floorplanning, their problem formulations and solution strategies are closely related to the pin assignment stage considered in this work. Therefore, this section reviews prior studies from these two perspectives.

\textbf{Package floorplanning:}
Existing studies on package floorplanning mainly focus on determining multi-chip locations, optimizing chip orientations, and reducing package-level interconnect cost. Liu et al.~\cite{liu2014floorplanning} used sequence pairs to represent the relative positions of dies, enumerated legal die locations, and selected the optimal layout based on HPWL to reduce the total wirelength at the interposer/PCB level. Seemuth et al.~\cite{seemuth2015automatic} applied simulated annealing to die placement and flexible I/O pin assignment, and further optimized pin assignment using ILP, effectively reducing wirelength. Seemuth et al.~\cite{seemuth2017flexible} reduced the required number of interposer layers through pin assignment. Starting from an initial pin assignment, they performed ILP-based routing to identify congested regions and then reassigned nets in those regions until no congestion remained. For EMIB packaging, Lee et al.~\cite{lee2023floorplanning} combined TCG and B*-tree into a hybrid layout representation and solved the resulting problem using simulated annealing, thereby optimizing package wirelength while satisfying EMIB constraints. Lin et al.~\cite{lin2023routability} proposed a routability-driven and orientation-aware placement method. They indirectly modeled legal chip orientations by introducing a continuous orientation variable and mapping it onto the four legal orientations through a softmax-based formulation during analytical optimization. They also constructed a congestion model supporting $0^\circ$, $90^\circ$, $180^\circ$, and $270^\circ$ routing, enabling simultaneous optimization of wirelength and routability.

Existing package floorplanning methods mainly focus on wirelength optimization or general routability modeling, but they do not sufficiently consider how the fan-out routing style affects congestion distribution in FOWLP. As a result, the congestion estimated during floorplanning may deviate from the actual routing bottlenecks encountered in subsequent RDL routing.

\textbf{Pin assignment:}
For pin assignment, related studies usually formulate the problem as endpoint mapping and propose exact or heuristic methods for different design scenarios. Kuo et al.~\cite{kuo2018pin} proposed an SLR-aware pin assignment method for multi-2.5D FPGA systems. Their method uses ILP-based iterative optimization to assign cross-FPGA signals to appropriate physical wires and SLR combinations, thereby improving subsequent routing and timing quality. Li et al.~\cite{li2016pin} formulated P/G pin assignment in large-scale packages as a multi-objective combinatorial optimization problem considering both PI and SI. They used customized simulated annealing to search the legal pin-swap space and combined objective normalization, initial-solution optimization, and static-template acceleration to generate high-quality pin assignment solutions within a short runtime. Zhuang et al.~\cite{zhuang2026adaptive} performed pin assignment using an MIS-based method. In each iteration, candidate optimal pins are generated for unassigned nets, and conflicting candidate assignments are modeled as a conflict graph. A set of mutually non-conflicting nets is then selected for assignment first, while the remaining nets are reserved for further optimization in subsequent iterations.

Although existing pin assignment methods can effectively reduce wirelength, they usually do not consider potential net crossings and their impact on subsequent routing. For inter-chip signal nets, pin-assignment solutions with the same HPWL may lead to different routing quality, and solutions with many potential crossings often result in longer routed wirelength.
\section{Preliminaries}
In this section, we first define the terminology and notations used in this article. We then present the considered advanced packaging design rules. Finally, we formulate the routability-driven package floorplanning problem with pin assignment.

\subsection{Terminology and Notations} The terminology, notations, and corresponding descriptions used in this article are  as follows:
\begin{itemize}
    \item $C=\{c_i \mid 1 \leq i \leq |C|\}$: the set of chips.
    \item $P=\{p_i \mid 1 \leq i \leq |P|\}$: the set of I/O pads belonging to inter-chip routing nets.
    \item $N=\{n_i \mid 1 \leq i \leq |N|\}$: the set of nets, where the two terminals of each net are two I/O pads.
    \item $\Theta=\left\{\theta_j \mid \theta_j=\frac{j\pi}{2},\; j=0,1,2,3 \right\}$: the set of legal chip orientations.
    \item $w_c$: The minimum spacing required between any two chips.
    \item $w_p$: The minimum spacing required between each chip and the package boundary.
\end{itemize}

\subsection{Design Rules}
To ensure physical feasibility and manufacturability, the following advanced packaging design rules are considered.

\textbf{Minimum Spacing Rule:}
The chip-to-chip spacing and chip-to-boundary spacing must be no smaller than the technology-dependent parameters $w_c$ and $w_p$, respectively.

\textbf{Rotation Rule:} 
Each chip can only be rotated to an orientation selected from the legal orientation set $\Theta$.

\textbf{Signal Connection Rule:}
Each terminal of a signal net must be assigned to a candidate I/O pad on its corresponding chip. A terminal must not be assigned to a pad on another chip, and each physical I/O pad can be assigned to at most one net.

\textbf{Fan-Out Routing Style:}
Signal nets are not allowed to use fan-in regions as routing resources and must be routed through fan-out regions. Accordingly, routability estimation should explicitly distinguish fan-in regions from fan-out regions.

\subsection{Problem Formulation}

Based on the above definitions and design rules, \textbf{the routability-driven package floorplanning problem with pin assignment} is formulated as follows: 

\textit{Given a set of chips $C$, a set of candidate I/O pads $P$ for inter-chip nets, a netlist $N$, a fixed outline of RDL, along with design rules, the objective is to find a floorplanning and pin-assignment solution that improves the routability of subsequent RDL routing while minimizing inter-chip routing wirelength.}

\begin{figure}[t!]
	\centering
	\includegraphics[width=0.65\linewidth,keepaspectratio]{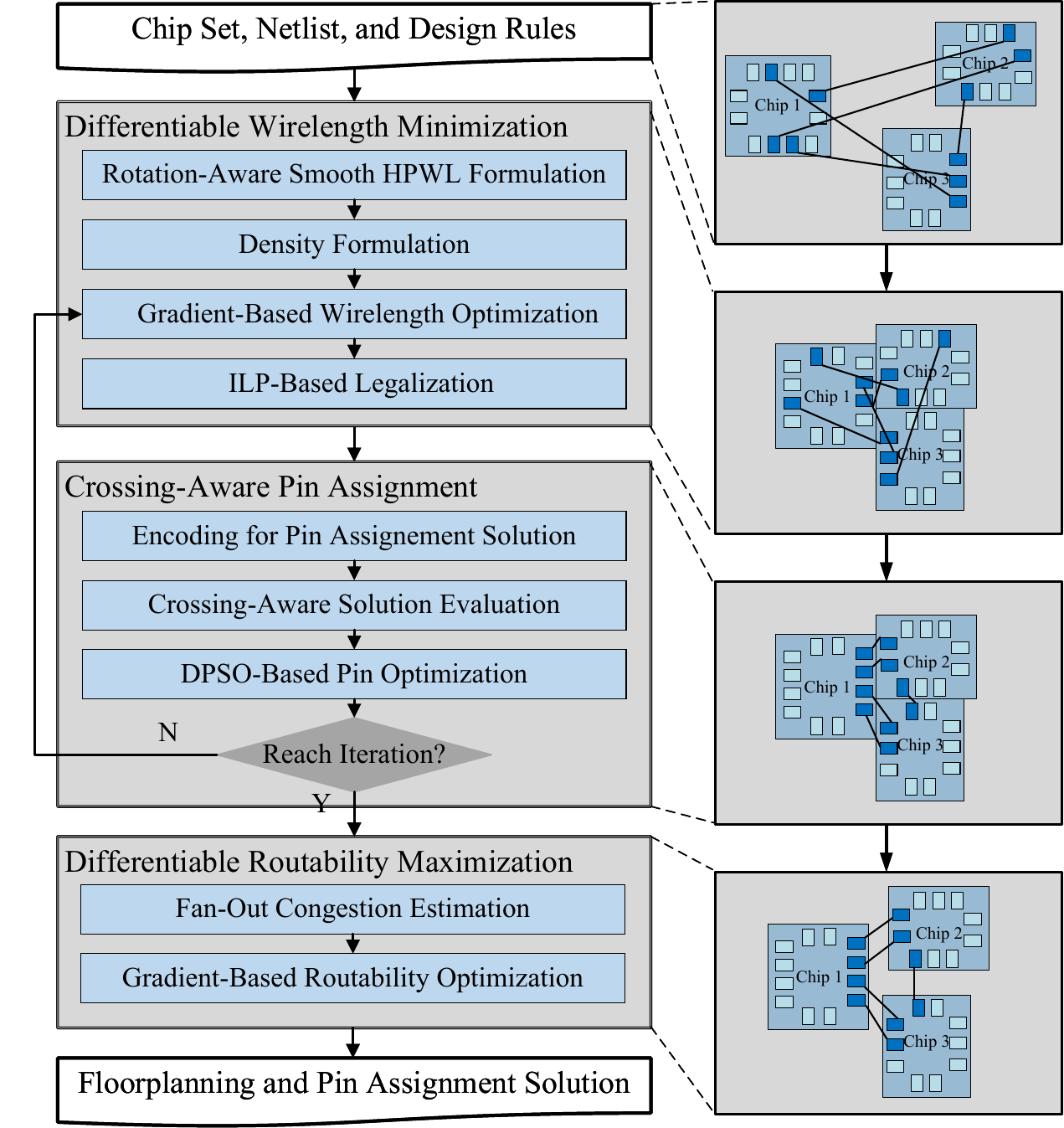}
	\caption{The flow of the proposed algorithm.}
	\label{f:4}
\end{figure}

\section{The Details of the Proposed Algorithm} 
Figure~\ref{f:4} illustrates the overall flow of the proposed differentiable routability-driven package floorplanning algorithm. The entire flow consists of three main stages, as follows.

\textbf{Differentiable Wirelength Minimization:}
This stage jointly optimizes chip locations and orientations to obtain a low-HPWL floorplanning solution, providing a high-quality initial layout for the subsequent differentiable routability maximization. First, a rotation-aware smooth HPWL model is constructed to capture the impact of chip orientations on I/O pad locations while maintaining differentiability. Then, gradient-based optimization is performed to update chip locations and orientation variables for wirelength reduction. Finally, an ILP-based legalization step is applied to remove overlaps and satisfy design constraints, producing a legal low-HPWL floorplanning solution for the next stage.

\textbf{Crossing-Aware Pin Assignment:}
Starting from the output of the previous stage, the crossing-aware pin assignment optimizes the pin selection for each net to reduce HPWL and potential net crossings. In the pin assignment encoding step, each pin assignment solution is represented as a particle. In the crossing-aware fitness evaluation step, a fitness function is defined to evaluate each particle by jointly considering HPWL and potential net-crossing cost. Based on this evaluation, the DPSO-based pin optimization step searches for a high-quality pin assignment solution. After this stage, the framework checks whether the specified iteration number has been reached. If not, it returns to the previous stage and continues the iterative optimization to further improve the floorplanning and pin assignment solution.

\textbf{Differentiable Routability Maximization:}
This stage takes the solution generated by the iterative execution of the previous two stages as input, which has short HPWL and few potential net crossings. Based on this solution, the stage expands the available spacing of critical routing channels through differentiable routability optimization, thereby improving RDL routability from an initial solution with short wirelength. In the fan-out congestion estimation, fan-out routing resources are extracted, and local routing demand is estimated using the Top-K candidate paths of signal nets. Based on this estimation, the subsequent gradient-based routability optimization converts the demand-capacity mismatch into a differentiable congestion cost, which is back-propagated to chip locations to guide chip movement and enlarge congested routing channels.

\subsection{Differentiable Wirelength Minimization}
To obtain an initial legal floorplanning solution with short interconnect length, the proposed framework first performs differentiable wirelength minimization. Since chip rotation changes the coordinates of I/O pads, the impact of chip orientation is explicitly incorporated into the wirelength estimation model, i.e., HPWL. However, the original two-terminal HPWL contains non-differentiable absolute-value operations and is therefore not suitable for gradient-based optimization. To address this issue, a smooth HPWL objective is constructed using the log-sum-exp approximation. In addition, a density regularization term is introduced to discourage chips from clustering during optimization. By incorporating smoothed HPWL and density regularization into a unified differentiable objective, the proposed formulation enables chip locations and orientations to be optimized through gradient-based updates. After optimization, an ILP-based legalization step is applied to enforce design constraints and generate a legal low-HPWL floorplanning solution for the subsequent stages.
\subsubsection{Rotation-Aware Smooth HPWL Formulation}
Let the center location of chip $i$ be $c_i=(x_i,y_i)$. The $j$-th pin of chip $i$ is denoted as $p_{i,j}$, and its local offset with respect to the chip center is given by $(\Delta x_j, \Delta y_j)$. When chip rotation is not considered, the global coordinate of pin $v$ can be written as
\begin{equation}
p_v = (x_i+\Delta x_j,\; y_i+\Delta y_j).
\label{eq:1}
\end{equation}

As shown in equation~\eqref{eq:1}, the coordinate of a pin is jointly determined by the chip center location and the local pin offset. We then consider chip rotation. When the rotation angle is $\theta_j \in \Theta$, the pin location should be rotated accordingly. Thus, the coordinate of pin $v$ under orientation $\theta_j$ is
\begin{equation}
p_v^{\theta_j}
=
\left(
x_i+\Delta x_v\cos\theta_j-\Delta y_v\sin\theta_j,\;
y_i+\Delta x_v\sin\theta_j+\Delta y_v\cos\theta_j
\right).
\label{eq:2}
\end{equation}

Since $\theta_j$ is discrete, it cannot be directly optimized by gradient-based algorithms. Therefore, for chip $i$, we introduce a learnable parameter $\alpha_{ij}$ for each legal orientation to indicate the tendency of chip $i$ to select orientation $\theta_j$. To obtain differentiable orientation-selection weights, Gumbel-Softmax~\cite{gumbel} is used to relax the discrete orientation selection into a continuous form. Let $g_{ij}\sim \mathrm{Gumbel}(0,1)$ denote the Gumbel noise. The relaxed weight for selecting orientation $\theta_j$ is defined as
\begin{equation}
	y_{ij}
	=
	\frac{\exp\left((\alpha_{ij}+g_{ij})/\tau\right)}
	{\sum_{k=0}^{3}\exp\left((\alpha_{ik}+g_{ik})/\tau\right)} ,
	\label{eq:3}
\end{equation}
where $\tau$ is the temperature parameter that controls the smoothness of orientation selection. When $\tau$ is large, the orientation weights are smoother, and multiple orientations may have comparable weights. As $\tau$ decreases, the relaxed orientation selection gradually approaches a one-hot form and tends to select a single legal orientation.
Based on the above orientation-selection weights, the relaxed coordinate of pin $v$ used for wirelength calculation is defined as
\begin{equation}
\tilde{p}_v
=
\sum_{j=0}^{3} y_{ij} p_v^{\theta_j}
=
(\tilde{x}_v,\tilde{y}_v).
\label{eq:4}
\end{equation}
This formulation avoids the dependence on the initialization of continuous rotation angles in indirect orientation modeling, leading to a more stable optimization process.

For wirelength estimation, HPWL is used to model wirelength in floorplanning. For net $n$, let the relaxed coordinates of its two terminals be $\tilde{p}_1=(\tilde{x}_1,\tilde{y}_1)$ and $\tilde{p}_2=(\tilde{x}_2,\tilde{y}_2)$. The HPWL of net $n$ is
\begin{equation}
	\mathrm{HPWL}
	=
	\left|\tilde{x}_1-\tilde{x}_2\right|
	+
	\left|\tilde{y}_1-\tilde{y}_2\right|.
	\label{eq:5}
\end{equation}

Although HPWL provides a simple wirelength estimate, the absolute-value operation in equation~\eqref{eq:5} is nondifferentiable when its input equals zero. Specifically, $\left|\tilde{x}_1-\tilde{x}_2\right|$ is non-differentiable when $\tilde{x}_1=\tilde{x}_2$, and the same problem also exists for the $y$-direction term. This may lead to unstable gradients during continuous floorplanning optimization. Therefore, the original HPWL is not suitable for direct use in gradient-based optimization. To improve differentiability, the log-sum-exp approximation is used to smooth the HPWL function. The smoothed HPWL of net $n$ is defined as
\begin{equation}
\begin{aligned}
\mathrm{HPWL}_{\mathrm{LSE}}(n)
=&\;
\gamma \log\left(e^{\tilde{x}_1/\gamma}+e^{\tilde{x}_2/\gamma}\right)
+
\gamma \log\left(e^{-\tilde{x}_1/\gamma}+e^{-\tilde{x}_2/\gamma}\right) \\
&+
\gamma \log\left(e^{\tilde{y}_1/\gamma}+e^{\tilde{y}_2/\gamma}\right)
+
\gamma \log\left(e^{-\tilde{y}_1/\gamma}+e^{-\tilde{y}_2/\gamma}\right),
\label{eq:6}
\end{aligned}
\end{equation}
where $\gamma$ is the smoothing parameter. A smaller $\gamma$ makes the approximation closer to the original HPWL but may cause sharper gradient changes. A larger $\gamma$ provides smoother gradients but increases the approximation error. Therefore, $\gamma$ should be properly set according to convergence stability and approximation accuracy. The HPWL loss over all signal nets is then obtained by accumulating the smoothed HPWL of each net:
\begin{equation}
	\mathcal{L}_{\mathrm{HPWL}}
	=
	\sum_{n\in N}
	\mathrm{HPWL}_{\mathrm{LSE}}(n),
	\label{eq:hpwl_loss}
\end{equation}

\subsubsection{Density Formulation}
Although $\mathcal{L}_{\mathrm{HPWL}}$ provides differentiable guidance for reducing inter-chip wirelength, optimizing wirelength alone may cause chips to cluster in a small local region of the package. Such clustering leads to an uneven spatial distribution and may require large displacements in the subsequent legalization step. Therefore, we introduce a differentiable density term as another objective in the wirelength minimization stage.

Let the package region be divided into $R\times C$ bins, and let $K=RC$ denote the total number of bins. The center of bin $k$ is denoted as $\mathbf{p}_k=(u_k,v_k)$. To compute the soft occupancy of chips over these bins, we first determine the axis-aligned footprint of each chip under the relaxed orientation.

It should be noted that chip rotation does not change the physical size of a chip. However, when a rectangular chip is rotated by $90^\circ$ or $270^\circ$, its width and height with respect to the global $x$- and $y$-axes are exchanged. Let the original physical size of chip $i$ be $w_i\times h_i$. The axis-aligned footprint width and height of chip $i$ under orientation $\theta_j$ are defined as
\begin{equation}
	(w_i^{\theta_j},h_i^{\theta_j})
	=
	\begin{cases}
		(w_i,h_i), & \theta_j=0^\circ \text{ or } 180^\circ,\\
		(h_i,w_i), & \theta_j=90^\circ \text{ or } 270^\circ.
	\end{cases}
	\label{eq:axis_aligned_footprint}
\end{equation}
Here, $w_i^{\theta_j}$ and $h_i^{\theta_j}$ denote the axis-aligned footprint width and height of chip $i$ under orientation $\theta_j$, rather than changes in the physical chip size.

Based on the orientation-selection weights $y_{ij}$ defined in equation~\eqref{eq:3}, the relaxed footprint width and height of chip $i$ are computed as
\begin{equation}
	\tilde{w}_i
	=
	\sum_{j=0}^{3}
	y_{ij}w_i^{\theta_j},
	\qquad
	\tilde{h}_i
	=
	\sum_{j=0}^{3}
	y_{ij}h_i^{\theta_j}.
	\label{eq:relaxed_chip_size}
\end{equation}
The relaxed axis-aligned boundaries of chip $i$ are then defined as
\begin{equation}
	l_i=x_i-\frac{\tilde{w}_i}{2},\quad
	r_i=x_i+\frac{\tilde{w}_i}{2},\quad
	b_i=y_i-\frac{\tilde{h}_i}{2},\quad
	t_i=y_i+\frac{\tilde{h}_i}{2}.
	\label{eq:relaxed_chip_boundary}
\end{equation}

Instead of using a hard binary indicator to determine whether bin $k$ is covered by chip $i$, we first compute one-dimensional soft inside scores along the $x$- and $y$-directions:
\begin{equation}
	\left\{
	\begin{aligned}
	s^x_{k,i}
	&=
	\operatorname{sigmoid}\left(\frac{u_k-l_i}{\tau_x}\right)
	\cdot
	\operatorname{sigmoid}\left(\frac{r_i-u_k}{\tau_x}\right),\\
	s^y_{k,i}
	&=
	\operatorname{sigmoid}\left(\frac{v_k-b_i}{\tau_y}\right)
	\cdot
	\operatorname{sigmoid}\left(\frac{t_i-v_k}{\tau_y}\right).
	\end{aligned}
	\right.
	\label{eq:soft_density_scores}
\end{equation}
where $\tau_x$ and $\tau_y$ are smoothing parameters. Here, $s^x_{k,i}$ measures whether the bin-center coordinate $u_k$ lies within the relaxed horizontal span $[l_i,r_i]$ of chip $i$, while $s^y_{k,i}$ measures whether $v_k$ lies within the relaxed vertical span $[b_i,t_i]$. If the bin center lies inside the corresponding span, the score approaches $1$; otherwise, it smoothly decreases toward $0$.

Therefore, the two-dimensional soft occupancy contribution of chip $i$ to bin $k$ is defined as
\begin{equation}
	o_{k,i}
	=
	s^x_{k,i}s^y_{k,i}.
	\label{eq:soft_occupancy}
\end{equation}
The total soft occupancy of bin $k$ is obtained by accumulating the contributions from all chips:
\begin{equation}
	o_k
	=
	\sum_i o_{k,i}.
	\label{eq:bin_occupancy}
\end{equation}

The density loss is defined as the variance of bin occupancies:
\begin{equation}
	\mathcal{L}_{\mathrm{density}}
	=
	\frac{1}{K}
	\sum_{k=1}^{K}
	(o_k-\bar{o})^2,
	\quad
	\bar{o}
	=
	\frac{1}{K}
	\sum_{k=1}^{K}o_k.
	\label{eq:density_loss}
\end{equation}
This term penalizes uneven chip distribution. When multiple chips are concentrated in a small region, some bins have much higher occupancy than others, increasing the occupancy variance and thus $\mathcal{L}_{\mathrm{density}}$. In contrast, when chips are distributed more uniformly across the package region, the bin occupancies become closer to their average value, reducing the density loss.

\subsubsection{Gradient-Based Wirelength Optimization}
The differentiable objective of this stage is defined by combining the smoothed HPWL term and the density regularization term:
\begin{equation}
	\mathcal{L}_1
	=
	\mathcal{L}_{\mathrm{HPWL}}
	+
	\lambda\mathcal{L}_{\mathrm{density}},
	\label{eq:wirelength_stage_loss}
\end{equation}
where $\lambda$ controls the relative importance of the density regularization term.
Based on the rotation-aware smooth HPWL formulation and the density formulation defined above, chip locations and orientations are jointly optimized by minimizing $\mathcal{L}_1$. The relaxed orientation weights $y_{ij}$ in equation~\eqref{eq:3} are used in two differentiable paths. First, they determine the relaxed pin coordinates in equation~\eqref{eq:4}, which are then used to compute the smoothed HPWL objective in equation~\eqref{eq:hpwl_loss}. Second, they determine the relaxed chip footprint in equation~\eqref{eq:relaxed_chip_size}, which further defines the relaxed chip boundaries and the soft bin occupancies used in the density loss in equation~\eqref{eq:density_loss}. Therefore, both terms in $\mathcal{L}_1$ can be back-propagated to the chip center locations $(x_i,y_i)$ and the orientation-selection parameters $\alpha_{ij}$, enabling gradient-based optimization of chip placement and orientation.

After continuous optimization, the orientation with the largest weight is selected as the legal orientation of each chip. For chip $i$, the orientation with the maximum weight is selected as its final orientation:
\begin{equation}
	\begin{cases}
		j_i^{}=\arg\max\limits_{j\in\{0,1,2,3\}} y_{ij},\\
		\theta_i^{\mathrm{final}}=\theta_{j_i^{}}.
	\end{cases}
\end{equation}
Since the final orientation is selected from the legal orientation set $\Theta$, the output orientation always satisfies the discrete orientation constraint.

\subsubsection{ILP-based Legalization}
After gradient-based wirelength optimization, the obtained chip locations may still violate the spacing rule. Therefore, in this step, ILP-based legalization is applied to adjust the chip center locations. Let $(\hat{x}_i,\hat{y}_i)$ denote the optimized center location of chip $i$ obtained from gradient-based wirelength optimization, and let $(x_i,y_i)$ denote its legalized center location. The objective of ILP-based legalization is to minimize the total displacement between the optimized and legalized chip locations:
\begin{equation}
	\min \sum_i \left(d_i^x+d_i^y\right),
	\label{eq:legalization_objective}
\end{equation}
subject to the following constraints.

First, the auxiliary variables $d_i^x$ and $d_i^y$ linearize the absolute displacement in the $x$ and $y$ directions, respectively:
\begin{equation}
	\begin{cases}
		d_i^x \ge x_i-\hat{x}_i, \\
		d_i^x \ge \hat{x}_i-x_i, \\
		d_i^y \ge y_i-\hat{y}_i, \\
		d_i^y \ge \hat{y}_i-y_i .
	\end{cases}
	\label{eq:legalization_displacement}
\end{equation}

Second, the ILP uses spacing-inflated rectangles to encode the minimum chip-to-chip spacing rule. Let $w_i^{final}$ and $h_i^{final}$ denote the width and height of chip $i$ under its final discrete orientation $\theta_i^{\mathrm{final}}$. The inflated dimensions used in the ILP are defined as
\begin{equation}
	w_i=w_i^{final}+w_c,\qquad h_i=h_i^{final}+w_c,
	\label{eq:inflated_dimensions}
\end{equation}
where $w_c$ denotes the chip-size inflation amount derived from the chip-to-chip spacing rule. Each spacing-inflated chip rectangle must be contained in the package outline while maintaining a package-boundary clearance $w_p$. Here, $W_{\mathrm{pkg}}$ and $H_{\mathrm{pkg}}$ denote the width and height of the package outline, respectively. The boundary constraints are formulated as
\begin{equation}
	\left\{
	\begin{aligned}
		w_p+\frac{w_i^{\mathrm{final}}}{2}
		&\le x_i \le
		W_{\mathrm{pkg}}-w_p-\frac{w_i^{\mathrm{final}}}{2}, \\
		w_p+\frac{h_i^{\mathrm{final}}}{2}
		&\le y_i \le
		H_{\mathrm{pkg}}-w_p-\frac{h_i^{\mathrm{final}}}{2}.
	\end{aligned}
	\right.
	\label{eq:boundary_constraints}
\end{equation}

Third, for any pair of spacing-inflated chip rectangles $i$ and $j$, at least one valid relative-position relationship must hold. Specifically, chip $i$ should be either to the left of, to the right of, below, or above chip $j$. To model this disjunctive constraint, four binary variables $l_{ij}$, $r_{ij}$, $b_{ij}$, and $t_{ij}$ are introduced. The non-overlap constraints are formulated as
\begin{equation}
	\begin{cases}
		x_i+\frac{w_i}{2} \le x_j-\frac{w_j}{2}+M(1-l_{ij}), \\
		x_j+\frac{w_j}{2} \le x_i-\frac{w_i}{2}+M(1-r_{ij}), \\
		y_i+\frac{h_i}{2} \le y_j-\frac{h_j}{2}+M(1-b_{ij}), \\
		y_j+\frac{h_j}{2} \le y_i-\frac{h_i}{2}+M(1-t_{ij}), \\
		l_{ij}+r_{ij}+b_{ij}+t_{ij} \ge 1, l_{ij},r_{ij},b_{ij},t_{ij}\in\{0,1\}.
	\end{cases}
	\label{eq:nonoverlap_constraints}
\end{equation}
where $M$ is a sufficiently large constant. Here, $l_{ij}=1$ indicates that chip $i$ is placed to the left of chip $j$, $r_{ij}=1$ indicates that chip $i$ is placed to the right of chip $j$, $b_{ij}=1$ indicates that chip $i$ is placed below chip $j$, and $t_{ij}=1$ indicates that chip $i$ is placed above chip $j$.

After solving the ILP, the legalized chip locations and fixed legal orientations form a floorplanning solution in which the spacing-inflated chip rectangles are non-overlapping and maintain the package-boundary clearance $w_p$. Therefore, the corresponding physical chips satisfy the original package-boundary, non-overlap, and minimum-spacing design rules. The legalized floorplanning solution is then passed to the crossing-aware pin assignment stage.

\subsection{Crossing-Aware Pin Assignment}
Based on the result obtained from the differentiable wirelength minimization stage, this stage further optimizes the pins at both ends of each signal net to reduce potential net crossings and the estimated wirelength, i.e., HPWL. To achieve this goal, we develop a DPSO-based pin assignment algorithm that incorporates multiple search strategies and is accelerated through GPU-based parallel computation.

The proposed algorithm begins with the encoding and evaluation of pin assignment solutions. Each feasible pin assignment is represented by a unique position code, which corresponds to a particle in the swarm, and its quality is measured by a fitness function. To avoid highly similar initial solutions, a uniform sampling strategy is employed to distribute particles across the solution space. During the iterative search, particles are divided into several subgroups according to their fitness values, and different update strategies are applied based on the relative quality of particles within each subgroup.
\subsubsection{Encoding for Pin Assignment Solutions}

In our problem, a package typically integrates multiple chips, and each chip contains a set of available pins for signal nets. For chip $x$, let $P_x=\{p_{x,1},p_{x,2},\ldots,p_{x,y_x}\}$ denote its ordered candidate-pin set, and let $N_x$ denote the set of signal nets that have one terminal on chip $x$. Since each terminal must be assigned to exactly one physical pin on its corresponding chip, the pin assignment problem can be represented by a chip-wise discrete vector. Therefore, a pin assignment solution $p_i$ can be encoded as
\begin{equation}
		p_i =
		\big(
		c(1,1), c(1,2), \cdots, c(1,y_1),
		\cdots,
		c(x,1), c(x,2), \cdots, c(x,y_x),
		\cdots,
		c(M,y_M)
		\big),
		\quad 1 \le x \le M.
	\label{eq:particle_position}
\end{equation}

For a multi-chip system containing $M$ chips, $c(x,k)=n$ indicates that the $k$-th candidate pin on chip $x$ is assigned to net $n$, while $c(x,k)=0$ indicates that this candidate pin is not selected by any signal net. The value of each component is constrained by
\begin{equation}
	c(x,k)\in N_x \cup \{0\},
	\quad 1\le x\le M,\; 1\le k\le y_x .
	\label{eq:pin_assignment_domain}
\end{equation}
Moreover, each net terminal on a chip must appear once and only once in the vector of that chip:
\begin{equation}
	\sum_{k=1}^{y_x}\mathbb{I}\big(c(x,k)=n\big)=1,
	\quad \forall n\in N_x,\; 1\le x\le M .
	\label{eq:pin_assignment_legality}
\end{equation}
This constraint prevents two pins on the same chip from being assigned to the same signal net, while also ensuring that every required net terminal is assigned to one feasible candidate pin.

\begin{figure}[t!]
	\centering
	\begin{minipage}[c]{0.76\linewidth}
		\centering
		\includegraphics[width=\linewidth,keepaspectratio]{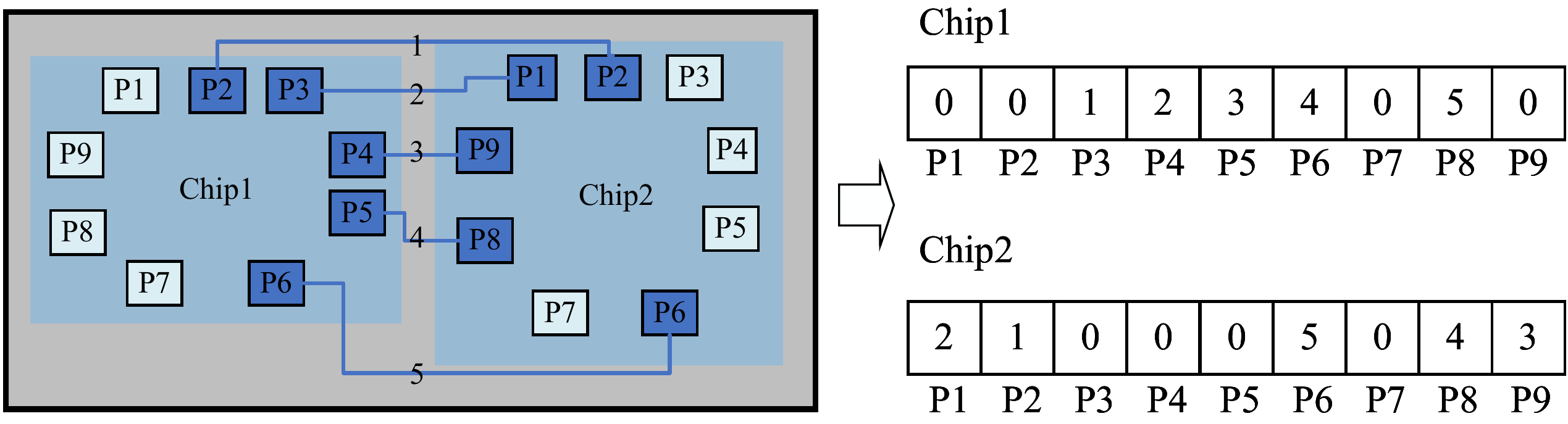}
	\end{minipage}
	\hspace{0.015\linewidth}
	\begin{minipage}[c]{0.1\linewidth}
		\centering
		\includegraphics[width=\linewidth,keepaspectratio]{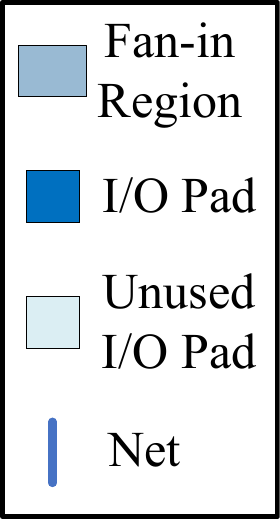}
	\end{minipage}
	\caption{Encoding of a pin assignment solution. Each chip is represented by an ordered pin vector, where a filled entry stores the net assigned to the corresponding pin.}
	\label{f:encoding}
\end{figure}

Figure~\ref{f:encoding} illustrates the proposed positional encoding with a simple example. For Chip 1, Nets 1--5 are assigned to pins $P3$, $P4$, $P5$, $P6$, and $P8$, respectively. For Chip 2, the same nets are assigned to pins $P2$, $P1$, $P9$, $P8$, and $P6$, respectively. The remaining pins on both chips are left unused and set to 0. The complete particle is then obtained by concatenating the ordered assignment vectors of all chips. With this representation, each particle corresponds to a feasible pin assignment candidate.

\subsubsection{Crossing-Aware Solution Evaluation}
Next, we define the evaluation metric for pin assignment solutions. Given fixed chip locations and orientations, a pin assignment determines the physical coordinates of all selected net terminals. A high-quality assignment should keep the estimated inter-chip wirelength short while avoiding pin configurations that are likely to cause routing competition in later RDL routing. The former effect can be captured by HPWL. For the latter effect, invoking a detailed RDL router for every particle during DPSO optimization would be prohibitively expensive. Therefore, we use flightline collisions as a lightweight early-stage routability proxy.

A flightline is defined as the direct line segment connecting the two assigned pins of a net, and a flightline collision occurs when the flightlines of two different nets intersect. Such a collision does not directly represent a detailed routing violation, because the final RDL router may resolve intersecting connections through layer assignment or detours. However, a large number of flightline collisions indicates that many nets are likely to compete for similar routing regions or require additional detours in subsequent RDL routing. Therefore, flightline collisions provide an early-stage routability indication for evaluating pin assignment solutions before detailed routing is performed. Since flightline collisions can be computed directly from assigned pin coordinates, they are also suitable for efficient fitness evaluation during DPSO-based pin optimization.

Based on the above analysis, the evaluation function of the pin assignment solution $p_i$ is defined as
\begin{equation}
	F(p_i) =
	\frac{
		\alpha\,\widehat{\mathrm{HPWL}}(p_i)
		+
		\beta\,\widehat{\mathrm{Crossing}}(p_i)
	}{T_{\mathrm{ref}}}.
	\label{eq:particle_fitness}
\end{equation}
where $\alpha$ and $\beta$ are two weighting factors, $\widehat{\mathrm{HPWL}}(p_i)$ denotes the normalized half-perimeter wirelength of particle $p_i$, and $\widehat{\mathrm{Crossing}}(p_i)$ denotes the normalized number of flightline collisions of particle $p_i$. In the implementation, the raw HPWL sum $H(p_0)$ and the raw flightline collision count $C(p_0)$ of the initial particle $p_0$ are used as reference scales:

\begin{equation}
	\widehat{\mathrm{HPWL}}(p_i)
	=
	\frac{H(p_i)}{H(p_0)},
	\quad
	\widehat{\mathrm{Crossing}}(p_i)
	=
	\frac{C(p_i)}{C(p_0)}.
	\label{eq:particle_cost_normalization}
\end{equation}
The normalization factor $T_{\mathrm{ref}}$ is the weighted total of the initial particle after applying the same normalization:
\begin{equation}
	T_{\mathrm{ref}}
	=
	\alpha\,\widehat{\mathrm{HPWL}}(p_0)
	+
	\beta\,\widehat{\mathrm{Crossing}}(p_0).
	\label{eq:particle_total_ref}
\end{equation}
The two terms are normalized before weighted summation to avoid dominance by either metric due to their different numerical scales. Since the optimization objective is to reduce both HPWL and crossings, a particle with a smaller fitness value represents a better pin assignment solution.

\begin{algorithm}[t]
	\caption{GPU-Parallel Warp-Level Evaluation of Particle Cost}
	\label{alg:gpu-particle-cost}
	\KwIn{particle swarm $\mathcal{P}$, number of nets $N$, warp size $w$}
	\KwOut{particle swarm costs $F(\mathcal{P})$}
	
	\ForEach{particle $p_m \in \mathcal{P}$ \textbf{in parallel}}{
		
		\ForEach{thread $\ell \in \{0,1,\ldots,w-1\}$ \textbf{in parallel}}{
			$h_{\ell}\gets 0$,\quad
			$c_{\ell}\gets 0$\;
			\For{$i = \ell+1$ to $N$ with step $w$}{
				$(a_1,a_2)\gets \operatorname{GetEndpoints}_{p_m}(i)$\;
				$h_{\ell}\gets h_{\ell}+\operatorname{CalHPWL}(a_1,a_2)$\;
				
				\For{$j = i+1$ to $N$}{
					$(b_1,b_2)\gets \operatorname{GetEndpoints}_{p_m}(j)$\;
					$c_{\ell}\gets c_{\ell}
					+\mathbb{I}\big(\operatorname{IsIntersect}(a_1,a_2,b_1,b_2)\big)$\;
				}
			}
		}
		
		$H_m\gets \operatorname{WarpReduceSum}(\{h_{\ell}\}_{\ell=0}^{w-1})$\;
		$C_m\gets \operatorname{WarpReduceSum}(\{c_{\ell}\}_{\ell=0}^{w-1})$\;
		
		$F(p_m)\gets \operatorname{CalCost}(H_m,C_m)$\;
	}
	\Return $F(\mathcal{P})=\{F(p_m)\mid p_m\in\mathcal{P}\}$\;
\end{algorithm}

In the subsequent DPSO-based pin optimization, the fitness values of a large number of particles need to be evaluated in each iteration. For each particle, both HPWL computation and flightline-collision evaluation involve a large number of independent operations over signal nets or flightline pairs. These operations can become a major computational bottleneck as the circuit scale increases. Therefore, it is necessary to accelerate this evaluation process. In this work, we introduce GPU-based parallel computation to evaluate both terms in the fitness function, enabling efficient evaluation of a large number of particles during the optimization process.

Algorithm~\ref{alg:gpu-particle-cost} shows the warp-level evaluation procedure for each particle. Each CUDA block contains one warp of 32 threads and is assigned to one particle, so the threads in the warp cooperatively evaluate the cost terms of that particle. For the HPWL term, each thread processes a strided subset of nets, obtains the two assigned endpoints of each net under the current particle assignment, and accumulates the HPWL contribution locally. For the flightline collision term, the same strided net traversal is used, while each selected net is compared only with nets of larger indices. In this way, only the upper-triangular set of net pairs is checked, avoiding repeated collision counting. After the per-thread partial sums are computed, warp-level reduction is applied to obtain the total HPWL term $H_m$ and the total flightline collision count $C_m$ of particle $m$. Finally, one thread writes the reduced values to global memory and evaluates the particle cost using the reference scales defined above.

\subsubsection{DPSO-Based Pin Optimization}
After encoding pin assignment solutions as particles and defining the corresponding evaluation function, we perform DPSO-based pin optimization to search for a high-quality pin assignment solution. Algorithm~\ref{alg:dpso-pin-optimization} summarizes the overall flow of the proposed optimization procedure. After particle initialization, the DPSO loop iteratively performs particle updates and best-solution updates until convergence.

\begin{algorithm}[t]
	\caption{DPSO-Based Pin Optimization}
	\label{alg:dpso-pin-optimization}
	\KwIn{particle swarm $\mathcal{P}$, maximum iteration number $T_{\max}$}
	\KwOut{Best pin assignment solution $\mathrm{Gbest}$}
	
	$P^0 \gets \operatorname{ParticleInitialization}()$\;
	$F(P^0) \gets \operatorname{EvaluationOfParticleCost}(P^0)$\;
	$R^0 \gets \operatorname{ReferenceParticleSelection}(P^0,F(P^0))$\;
	
	\For{$t= 0$ to $T_{\max}$}{
		\If{convergence}{\textbf{break}\;}
		$P^{t+1} \gets \operatorname{ParticleUpdate}(P^t)$\;
		
		$F(P^{t+1}) \gets \operatorname{EvaluationOfParticleCost}(P^{t+1})$\;
		$R^{t+1} \gets \operatorname{ReferenceParticleSelection}(P^{t+1},F(P^{t+1}))$\;
	}
	
	\Return $\mathrm{Gbest}\in R^t$\;
\end{algorithm}

\begin{figure}[t]
	\centering
	\includegraphics[width=0.85\linewidth,keepaspectratio]{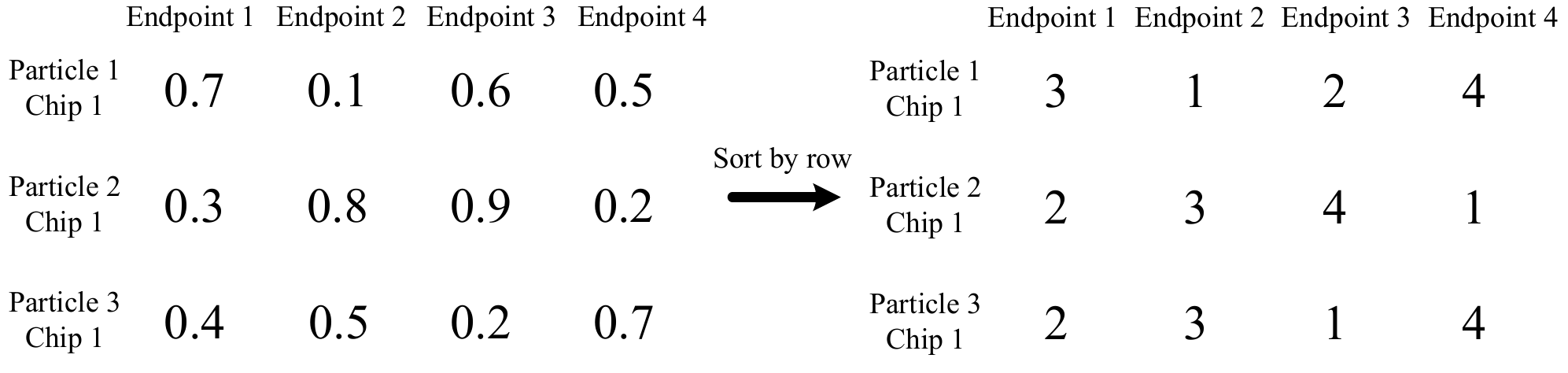}
	\caption{Stratified uniform initialization of pin assignment particles using sorted random keys.}
	\label{f:initialization}
\end{figure}

\text{1) Particle Initialization: }Before the formal search process starts, the initial particles are generated by a stratified uniform sampling strategy. As shown in Figure~\ref{f:initialization}, for each chip, we consider the net endpoints assigned to this chip across all particles and construct a sampling matrix, where each row corresponds to one particle and each column corresponds to one endpoint dimension. For each dimension, the interval $[0,1]$ is uniformly divided into $N_p$ non-overlapping subintervals according to the number of particles, and one random value is sampled from each subinterval. In this way, the samples of each endpoint dimension are spread over the entire interval rather than concentrated in a small region. After the sampling values of all dimensions are generated, the values in the same row are combined to form the random-key vector of one particle. The endpoints are sorted according to their random-key values to obtain an endpoint ordering. Then, $N_E$ pads are randomly selected from the candidate pad set of the chip to form a pad sequence, where $N_E$ is the number of endpoints on this chip. The sorted endpoints are assigned to the selected pads in this sequence, producing a legal chip-wise pin assignment. Since each selected pad is used at most once and each endpoint appears exactly once, the generated particle satisfies the pin assignment legality constraint.

\text{2) Particle Update: } To maintain particle diversity during the search process, the proposed algorithm partitions the swarm into four subgroups according to particle fitness values and applies differentiated particle update strategy to different subgroups. This design enhances the search for high-quality pin-assignment solutions while reducing the risk of premature convergence to local optima.

At iteration $t$, let $N_p$ denote the swarm size. Since a smaller fitness value indicates a better pin assignment solution, all particles are first sorted in ascending order of fitness:
\begin{equation}
	F(p_{\pi_1}^t)
	\le
	F(p_{\pi_2}^t)
	\le
	\cdots
	\le
	F(p_{\pi_{N_p}}^t),
	\label{eq:particle_sorting}
\end{equation}
where $\pi_k$ denotes the index of the $k$-th particle in the sorted order. The sorted particle sequence is then divided into four consecutive subgroups with nearly equal sizes:
\begin{equation}
	\begin{aligned}
	\mathcal{G}_j^t =
	\left\{
	p_{\pi_k}^t \mid b_j \le k \le e_j
	\right\},
	\quad
	b_j = \left\lfloor \frac{(j-1)N_p}{4} \right\rfloor + 1,\quad
	e_j = \left\lfloor \frac{jN_p}{4} \right\rfloor,
	\quad j=1,2,3,4 .
	\end{aligned}
	\label{eq:particle_subgroup}
\end{equation}
This sorted partition ensures that each subgroup is non-empty when $N_p \ge 4$, and the subgroup sizes differ by at most one. The subgroups are ordered from high-quality to low-quality particles; thus, $\mathcal{G}_1^t$ contains the best particles and $\mathcal{G}_4^t$ contains the worst particles. Accordingly, $\mathrm{Sbest}_1^t$ is identical to $\mathrm{Gbest}^t$.

\begin{figure*}[t]
	\centering
	\includegraphics[width=0.5\linewidth,keepaspectratio]{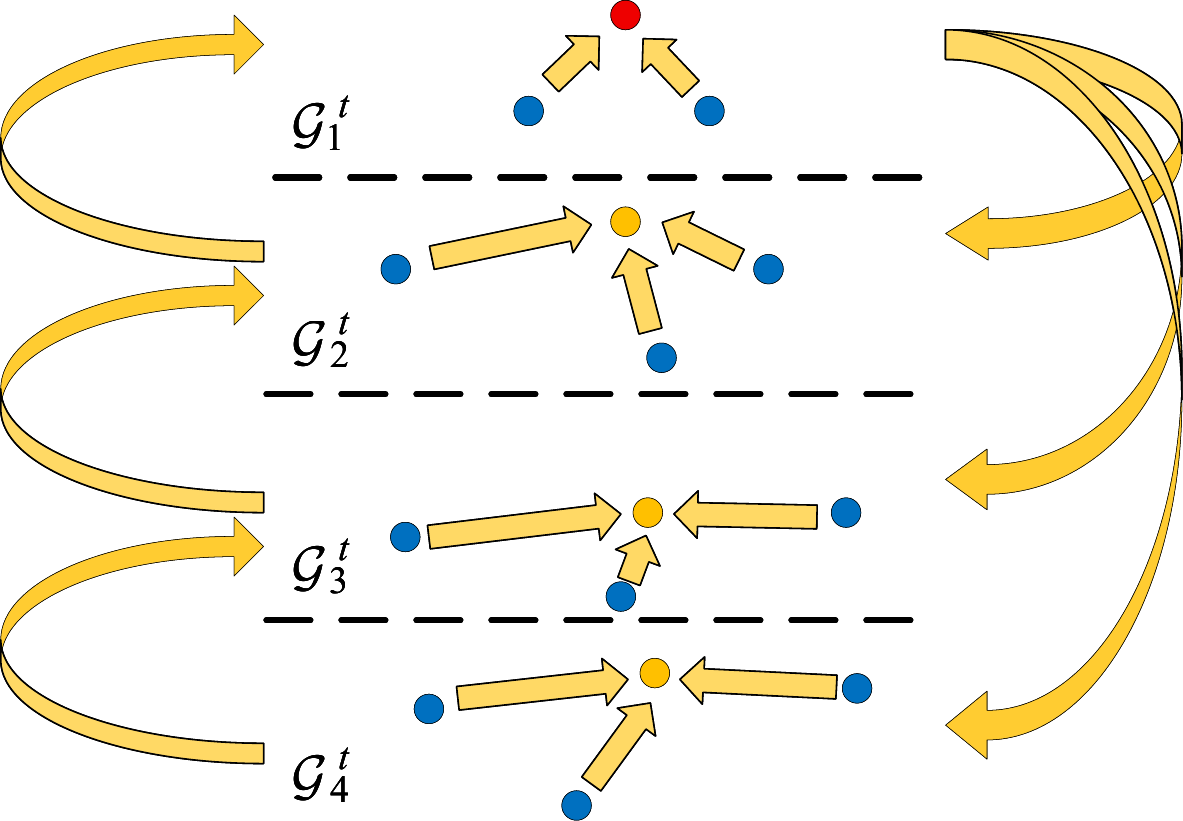}
	\vspace{3pt}
	\includegraphics[width=0.55\linewidth,keepaspectratio]{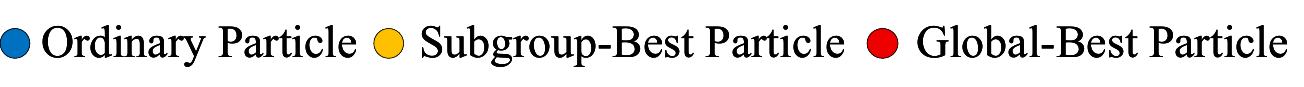}
	\caption{Subgroup partitioning and differential particle update strategy.}
	\label{f:swarm}
\end{figure*}

Based on the subgroup partitioning result, the proposed method updates particles with different roles using differentiated mechanisms. Specifically, particles are classified into three types: the global-best particle, non-global subgroup-best particles, and ordinary particles. As shown in Figure~\ref{f:swarm}, the global-best particle provides the overall search guidance and selects a reference particle from the non-global subgroup-best particles using roulette-wheel selection. Each non-global subgroup-best particle propagates high-quality assignment information within its own subgroup and learns from the adjacent higher-quality subgroup to improve its search direction. Ordinary particles mainly learn from their personal best solutions and the best particle in their corresponding subgroup. Through this role-aware update mechanism, the swarm can strengthen information exchange among particles, maintain search diversity, and improve the probability of escaping local optima. The update of particle $p_i^t$ can be generally expressed as
\begin{equation}
	p_i^{t+1}
	=
	F_3\Big(
	F_2\big(
	F_1(p_i^t,\omega_i^t),
	c_{11},c_{12},c_{13}
	\big),
	c_{21},c_{22},c_{23}
	\Big).
	\label{eq:particle_update}
\end{equation}
where $\omega_i^t$ denotes the inertia weight of the $i$-th particle at iteration $t$, which determines the probability of performing the mutation operation. 
$c_{1x}$ and $c_{2x}$ $(x=1,2,3)$ are acceleration coefficients that control the crossover probabilities for the three different types of particles.
$F_1$ represents the inertial component, which preserves the current search state of the particle. 
$F_2$ represents the individual cognitive component, which guides the particle to learn from its own historical best position. 
$F_3$ represents the social cognitive component, which encourages the particle to learn from elite particles in its current subgroup or from higher-quality subgroup-best particles.
The inertial component of a particle is denoted by $F_1$, which is implemented through a mutation operation. It is formulated as follows:
\begin{equation}
	W_i^t
	=
	F_1\left(p_i^t, \omega_i^t\right)
	=
	\begin{cases}
		\operatorname{swap}(c,x_a,x_b), 
		& r_1 < \omega_i^t, \\[2pt]
		\operatorname{rep}(c,x_a), 
		& r_1 \ge \omega_i^t \wedge P_c^{\mathrm{unused}} \ne \varnothing, \\[2pt]
		p_i^t, 
		& \text{otherwise}.
	\end{cases}
	\label{eq:inertial_component}
\end{equation}

where $W_i^t$ denotes the particle after the inertial update, and $r_1$ is a random number uniformly sampled from $[0,1)$. 
$P_c^{\mathrm{unused}}$ represents the set of unused candidate pads on chip $c$. 
For the current particle $p_i$, $\operatorname{swap}(c,x_a,x_b)$ updates the particle by exchanging the net endpoints assigned to two pads $x_a$ and $x_b$ within the same chip $c$. 
In contrast, $\operatorname{rep}(c,x_a)$ selects one unused candidate pad from $P_c^{\mathrm{unused}}$ and moves the net endpoint assigned to pad $x_a$ to that pad. 

\begin{figure*}[t]
	\centering
	
	\begin{subfigure}[b]{0.40\linewidth}
		\centering
		\includegraphics[width=\linewidth]{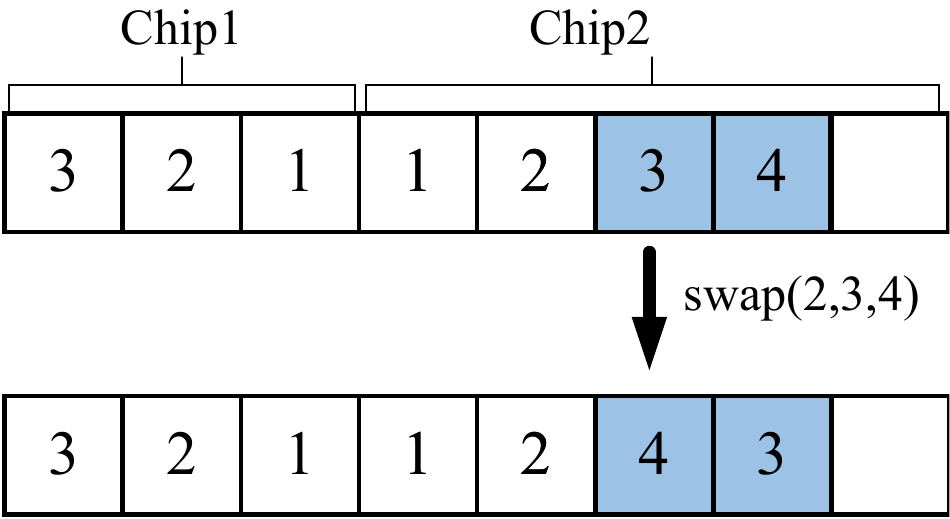}
		\caption{Swap operator.}
		\label{f:swap}
	\end{subfigure}
	\hspace{0.03\linewidth}
	\begin{subfigure}[b]{0.40\linewidth}
		\centering
		\includegraphics[width=\linewidth]{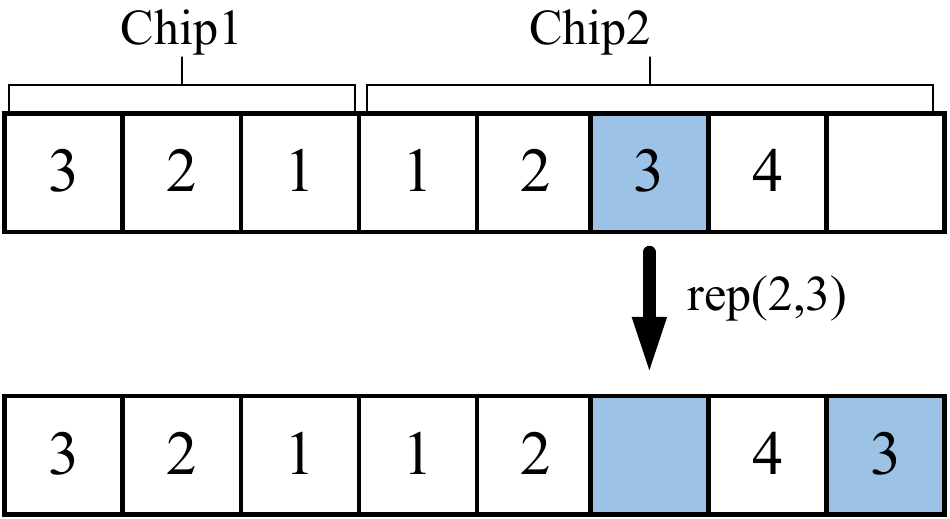}
		\caption{Replacement operator.}
		\label{f:rep}
	\end{subfigure}
	
	\caption{Examples of swap and replacement operators in chip-wise pin assignment.}
	\label{f:swap&&rep}
	
\end{figure*}

Figure~\ref{f:swap&&rep} illustrates the two discrete operators. In the swap operation shown in Figure~\ref{f:swap&&rep}(\subref{f:swap}), the assignments of two selected pads on the same chip are exchanged; for example, Nets 3 and 4 swap their assigned pads while the used-pad set remains unchanged. In the replacement operation shown in Figure~\ref{f:swap&&rep}(\subref{f:rep}), the net assigned to a selected pad is moved to an unused pad on the same chip; for example, Net 3 is reassigned to a new pad and the original pad becomes empty.
To balance the global and local exploitation ability of the algorithm, an adaptive inertia weight is further introduced. It is defined as follows:
\begin{equation}
	\omega_i^t
	=
	\begin{cases}
		\dfrac{\overline{F}\left(\mathrm{Sbest}^t\right)}
		{F\left(p_i^t\right)}, 
		& p_i^t = \mathrm{Gbest}^t, \\[8pt]
		\dfrac{F\left(\mathrm{Sbest}_{j-1}^t\right)}
		{F\left(p_i^t\right)}, 
		& p_i^t = \mathrm{Sbest}_j^t,\ j=2,3,4, \\[8pt]
		\dfrac{F\left(\mathrm{Sbest}_{S(i)}^t\right)}
		{F\left(p_i^t\right)}, 
		& \text{otherwise}.
	\end{cases}
	\label{eq:adaptive_inertia_weight}
\end{equation}

where $\mathrm{Gbest}^t$ denotes the global best particle at iteration $t$, and $\mathrm{Sbest}_j^t$ denotes the best particle in the $j$-th subgroup at iteration $t$. 
$F(p_i^t)$ represents the fitness value of the current particle. 
$\overline{F}(\mathrm{Sbest}^t)$ denotes the average fitness value of the best particles from all subgroups at iteration $t$. 
$F(\mathrm{Sbest}_{j-1}^t)$ represents the fitness value of the best particle in the adjacent higher-quality subgroup at iteration $t$. 
$F(\mathrm{Sbest}_{S(i)}^t)$ denotes the fitness value of the best particle in the subgroup to which the $i$-th particle belongs. 
With this strategy, each particle obtains an inertia weight according to the elite reference associated with its role: the global-best particle uses the average quality of subgroup-best particles, non-global subgroup-best particles use the adjacent higher-quality subgroup-best, and ordinary particles use the best particle in their own subgroup. Then, the inertia weight is further normalized as follows:
\begin{equation}
	\omega_i^{t*}
	=
	0.4 + 0.5 \cdot \Phi\left(\omega_i^t\right),
	\label{eq:normalized_inertia_weight}
\end{equation}

where $\Phi(\cdot)$ is the cumulative distribution function of the standard normal distribution. After this normalization, all inertia weights are mapped into the interval $[0.4,0.9]$, which is a commonly used range for inertia weights in PSO variants. 
In this way, different particles can adaptively adjust their update intensities according to their own search states, thereby improving the overall search efficiency and robustness of the algorithm.

The individual cognitive component of a particle is denoted by $F_2$, which is implemented through a crossover operation. It is formulated as follows:
\begin{equation}
	I_i^t
	=
	F_2\left(W_i^t,c_{1x}\right)
	=
	\begin{cases}
		C\left(W_i^t,p_i^{\mathrm{pbest}}\right),
		& r_2 < c_{1x}, \\[2pt]
		W_i^t,
		& \text{otherwise}.
	\end{cases}
	\label{eq:individual_cognitive_component}
\end{equation}

where $I_i^t$ denotes the particle after the individual cognitive update, and $r_2$ is a random number uniformly sampled from $[0,1)$. 
$c_{1x}$ $(x=1,2,3)$ is an acceleration coefficient that controls the probability of applying the crossover operation. 
Its value is determined according to the particle type, allowing different particles to have different crossover probabilities. 
$C(W_i^t,p_i^{\mathrm{pbest}})$ denotes the crossover operation between the particle after the inertial update $W_i^t$ and its historical best position $p_i^{\mathrm{pbest}}$, thereby guiding the particle to learn from its own best search experience.

\begin{figure}[t!]
	\centering
	\includegraphics[width=0.65\linewidth,keepaspectratio]{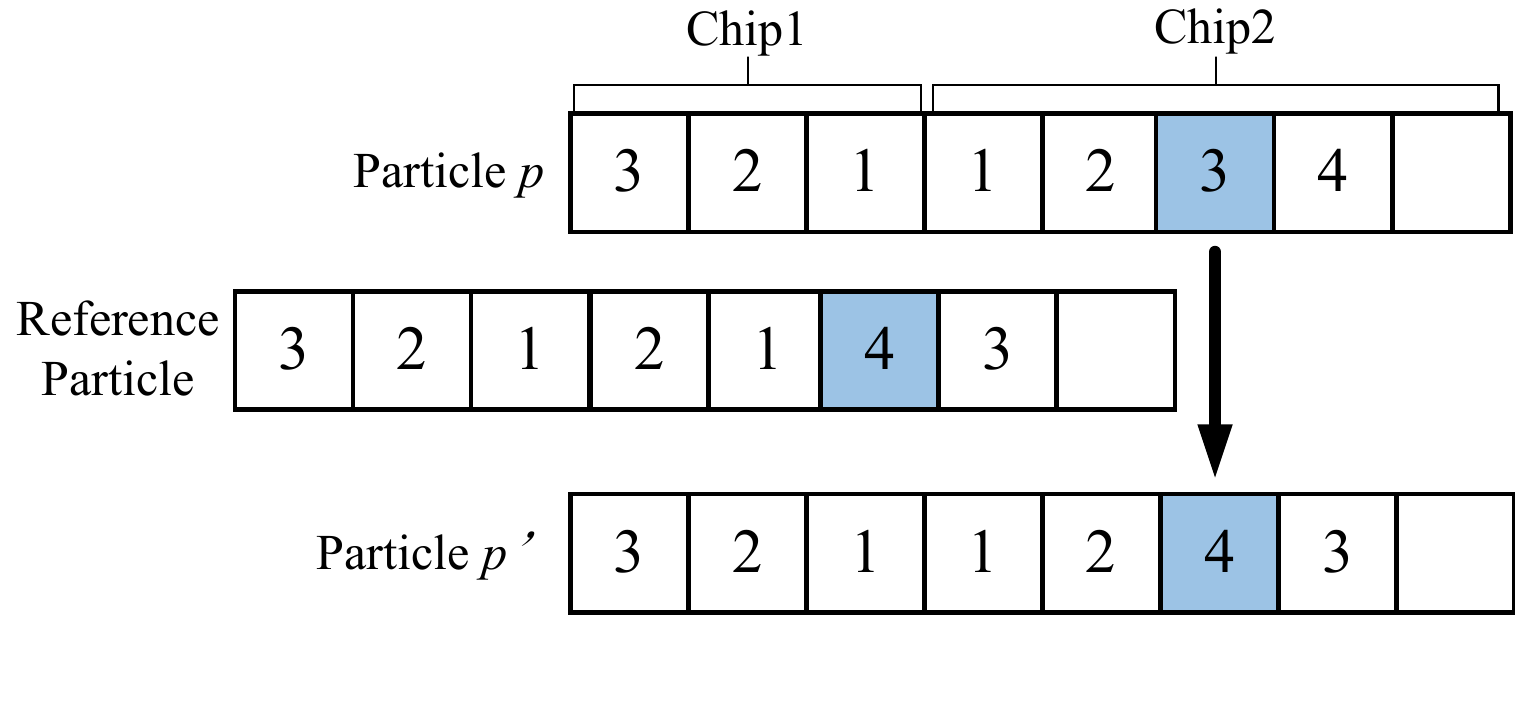}
	\caption{Discrete crossover operation for pin assignment particles.}
	\label{f:corssover}
\end{figure}

In the pin assignment problem, the crossover operation can be regarded as a discrete movement of a particle toward a reference particle. 
In the individual cognitive component, the reference particle is the individual historical best solution. 
Specifically, one position that is inconsistent with the reference particle is randomly selected, and the current particle is partially aligned with the reference particle through a swap operation. 
To preserve the legality of chip-wise pin assignments, this crossover operation is performed within the same chip-wise assignment vector.
Let the selected position be $X$, and let the value at the same position in the reference particle be $X_{\mathrm{ref}}$. 
Then, the position whose value is $X_{\mathrm{ref}}$ is searched for in the current particle and denoted as $Y$. 
By exchanging the values at positions $X$ and $Y$, the particle is moved one step closer to the reference particle. 
After each crossover operation, particles of different types approach their individual best solutions with different probabilities. 
Therefore, during the population evolution process, particles can preserve part of the promising assignment patterns from their historical best solutions.
Figure~\ref{f:corssover} illustrates an example of this crossover operation. In the selected chip-wise vector, the selected position is assigned to Net 3 in the current particle, while the corresponding position in the reference particle is assigned to Net 4. The crossover operation searches for Net 4 in the current particle and swaps the two assignments, so that the selected position is updated from Net 3 to Net 4.

The social cognitive component of a particle is denoted by $F_3$, which is also implemented through a crossover operation. It is formulated as follows:
\begin{equation}
	G_i^t
	=
	F_3\left(I_i^t,c_{2x}\right)
	=
	\begin{cases}
		C\left(I_i^t,p_i^{\mathrm{ref}}\right),
		& r_3 < c_{2x}, \\[2pt]
		I_i^t,
		& \text{otherwise}.
	\end{cases}
	\label{eq:social_cognitive_component}
\end{equation}

where $G_i^t$ denotes the particle after the social cognitive update, and $r_3$ is a random number uniformly sampled from $[0,1)$. 
$c_{2x}$ $(x=1,2,3)$ is a learning coefficient that controls the probability of applying the crossover operation. 
When $r_3 < c_{2x}$, particle $I_i^t$ performs the crossover operation with its reference particle $p_i^{\mathrm{ref}}$. 
In this work, particles are divided into three types, and different types of particles select different reference particles during the social cognitive update. 
The reference particle is selected as follows:
\begin{equation}
	p_i^{\mathrm{ref}}
	=
	\begin{cases}
		\operatorname{Roulette}\left(C^t\right),
		& p_i^t = \mathrm{Gbest}^t, \\[2pt]
		\mathrm{Sbest}_{j-1}^t,
		& p_i^t = \mathrm{Sbest}_j^t,\ j=2,3,4, \\[2pt]
		\mathrm{Sbest}_{S(i)}^t,
		& p_i^t = \mathrm{Ordinary}.
	\end{cases}
	\label{eq:reference_particle_selection}
\end{equation}

where
\begin{equation}
	C^t
	=
	\left\{
	\mathrm{Sbest}_k^t
	\mid
	\mathrm{Sbest}_k^t \ne \mathrm{Gbest}^t,\ 
	k=1,2,3,4
	\right\}
\end{equation}
denotes the set of subgroup-best particles excluding the global best particle. 
$\mathrm{Sbest}_{j-1}^t$ denotes the best particle in the adjacent higher-quality subgroup at iteration $t$, and $\mathrm{Sbest}_{S(i)}^t$ denotes the best particle in the subgroup to which particle $i$ belongs. 
That is, ordinary particles learn from the best particle in their own subgroup, non-global subgroup-best particles learn from the best particle in the adjacent higher-quality subgroup, and the global best particle selects its reference particle from the best particles of other subgroups using a roulette-wheel selection strategy. Specifically, for each candidate particle $q_k^t \in C^t$, its selection weight is constructed using the reciprocal of its fitness value:
\begin{equation}
	\eta_k^t
	=
	\frac{1}{F\left(q_k^t\right)} .
	\label{eq:roulette_weight}
\end{equation}

The corresponding selection probability is then obtained by normalization:
\begin{equation}
	\rho_k^t
	=
	\frac{\eta_k^t}
	{\sum_{q_j^t \in C^t} \eta_j^t}.
	\label{eq:roulette_probability}
\end{equation}

Next, let $R_k^t$ denote the cumulative selection probability of the $k$-th candidate particle at iteration $t$:
\begin{equation}
	R_k^t
	=
	\sum_{j=1}^{k} \rho_j^t,
	\label{eq:cumulative_probability}
\end{equation}
where $R_0^t=0$. 
A random number $r \sim U(0,1)$ is then generated. 
If
\begin{equation}
	R_{k-1}^t < r \le R_k^t,
	\label{eq:roulette_condition}
\end{equation}
then the $k$-th subgroup-best particle is selected as the reference particle, namely
\begin{equation}
	p_i^{\mathrm{ref}}
	=
	q_k^t .
	\label{eq:roulette_result}
\end{equation}

Through this strategy, the global best particle can absorb useful information from other high-quality subgroups while preserving its own advantageous structure, thereby enhancing population diversity and preventing premature convergence. In addition, the crossover operation used in the social cognitive component is the same as that in the individual cognitive component. By repeatedly performing crossover operations, particles can continuously learn promising assignment patterns from the population and gradually move toward the global best solution.

\subsection{Differentiable Routability Maximization}
\begin{figure*}[t]
	\centering
	\includegraphics[width=0.7\linewidth]{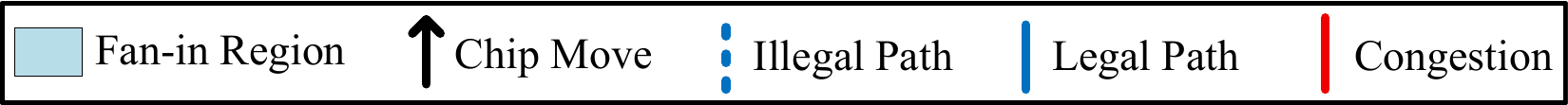}
	\vspace{1mm}
	\begin{subfigure}[b]{0.35\linewidth}
		\centering
		\includegraphics[width=\linewidth]{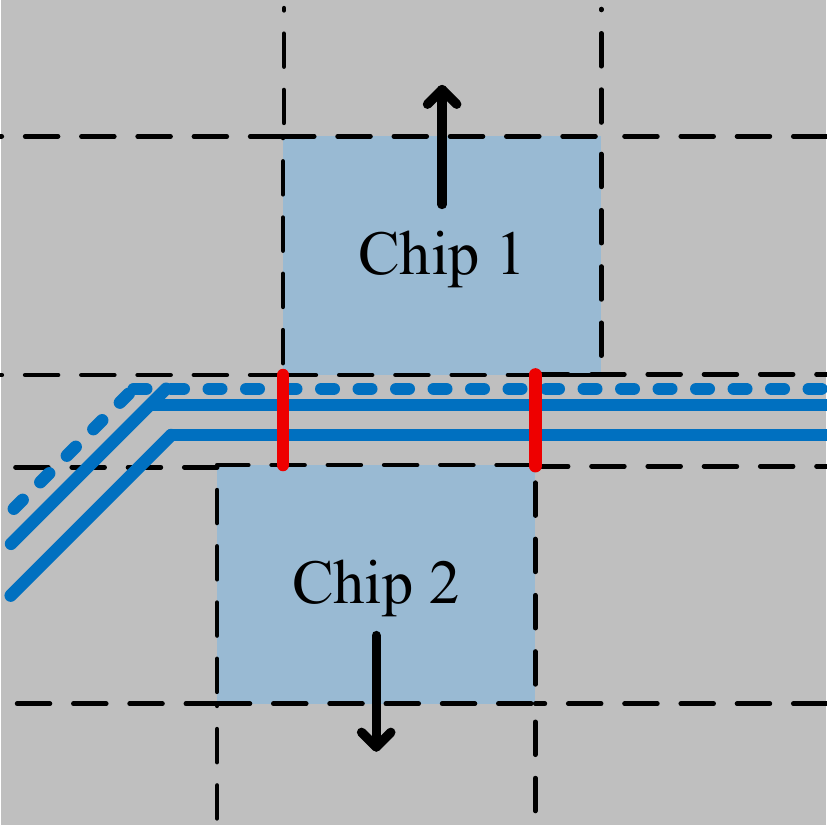}
		\caption{}
		\label{f:stage2a}
	\end{subfigure}
	\hspace{0.05\linewidth}
	\begin{subfigure}[b]{0.35\linewidth}
		\centering
		\includegraphics[width=\linewidth]{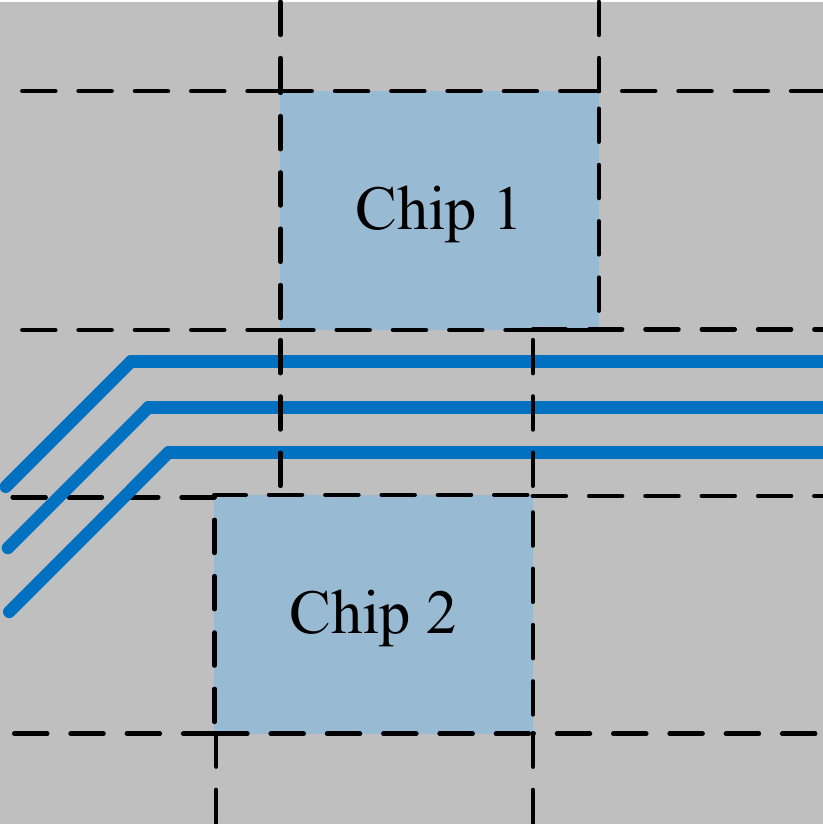}
		\caption{}
		\label{f:stage2b}
	\end{subfigure}
	\caption{(a) Congestion occurs in the routing channel between Chip 1 and Chip 2. (b) The differentiable optimization algorithm guides chip movement to alleviate the congestion.}
	\label{f:stage2}
\end{figure*}

In this stage, the proposed method improves routability through fan-out congestion estimation and gradient-based routability optimization. In fan-out congestion estimation, we model the actual usage of routing resources under the fan-out routing style by excluding fan-in regions from available routing resources and estimating the potential demand of routing channels. Based on this congestion model, the estimated congestion cost is incorporated into a differentiable optimization framework. In gradient-based routability optimization, we update chip positions according to the congestion gradients, guiding the layout toward lower congestion and higher routability. As shown in Figure~\ref{f:stage2}, congestion occurs in the routing channel between Chip 1 and Chip 2. Based on the congestion information, the algorithm guides the movement of chip locations to enlarge the congested routing channel, thereby alleviating the congestion. It should be noted that the routing graph construction and Top-$K$ path search are performed as discrete congestion-estimation steps. During each gradient-based optimization interval, the estimated routing demand $\mathrm{dem}(v)$ and the vertex-to-channel association are kept fixed, and the differentiable loss is back-propagated only through the routing channel length $l_v$, which is determined by chip boundaries and chip center locations. When a chip movement changes the routing-region topology, the routing graph and candidate paths are reconstructed to refresh the congestion estimation. 
\subsubsection{Fan-Out Congestion Estimation}
\begin{figure*}[t]
	\centering
	\includegraphics[width=1\linewidth]{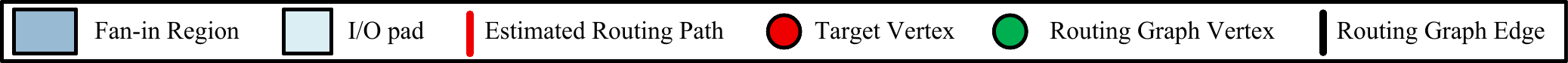}\par
	\vspace{1mm}
	\begin{subfigure}[b]{0.33\textwidth}
		\centering
		\includegraphics[width=\linewidth]{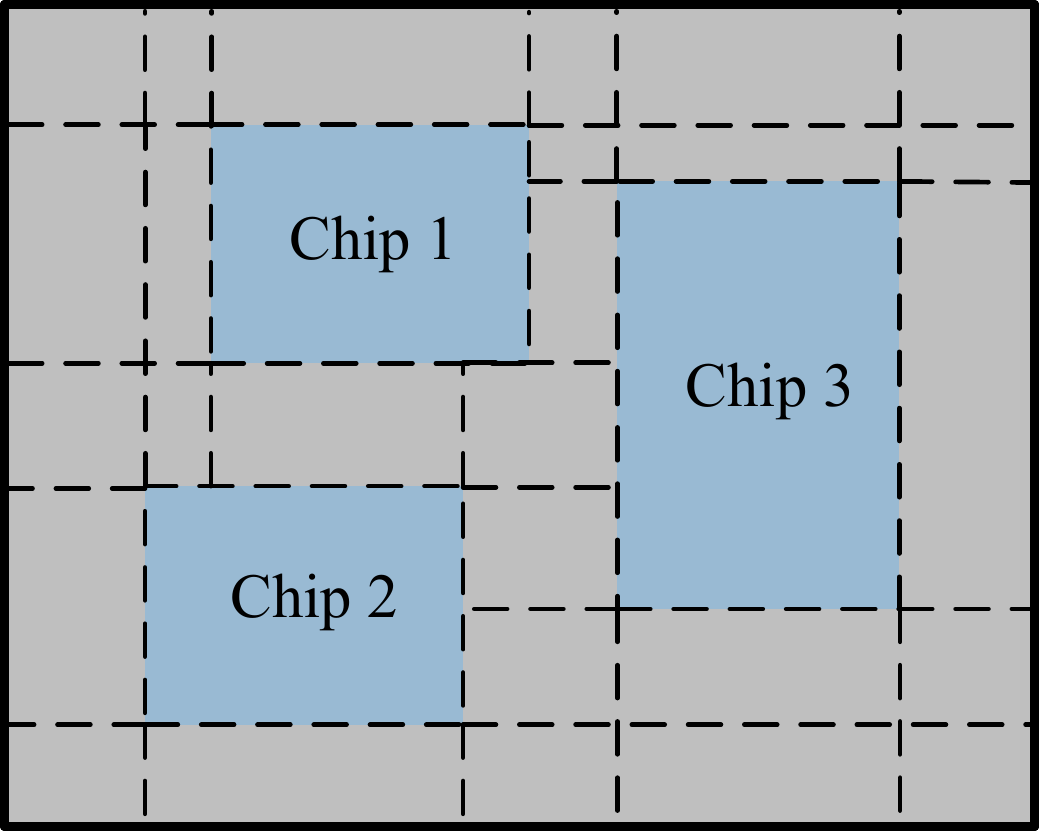}
		\caption{}
		\label{f:routingregion_a}
	\end{subfigure}
	\hfill
	\begin{subfigure}[b]{0.33\textwidth}
		\centering
		\includegraphics[width=\linewidth]{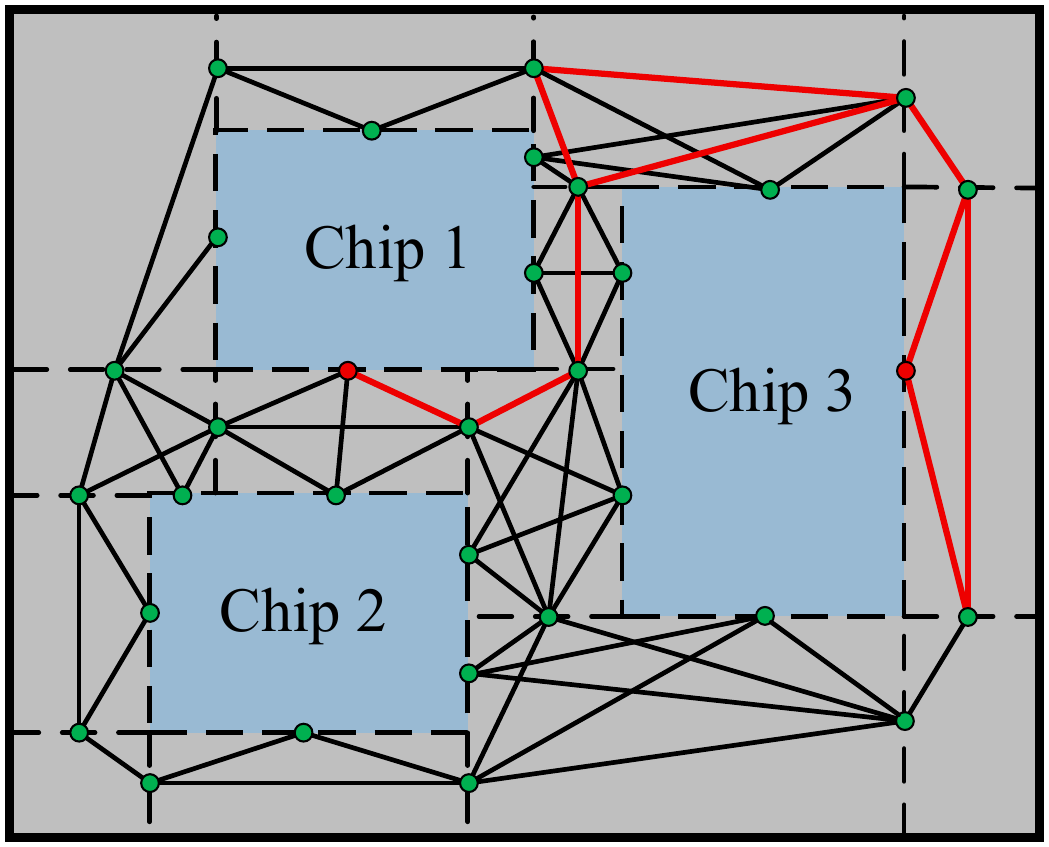}
		\caption{}
		\label{f:routingregion_b}
	\end{subfigure}
	\hfill
	\begin{subfigure}[b]{0.33\textwidth}
		\centering
		\includegraphics[width=\linewidth]{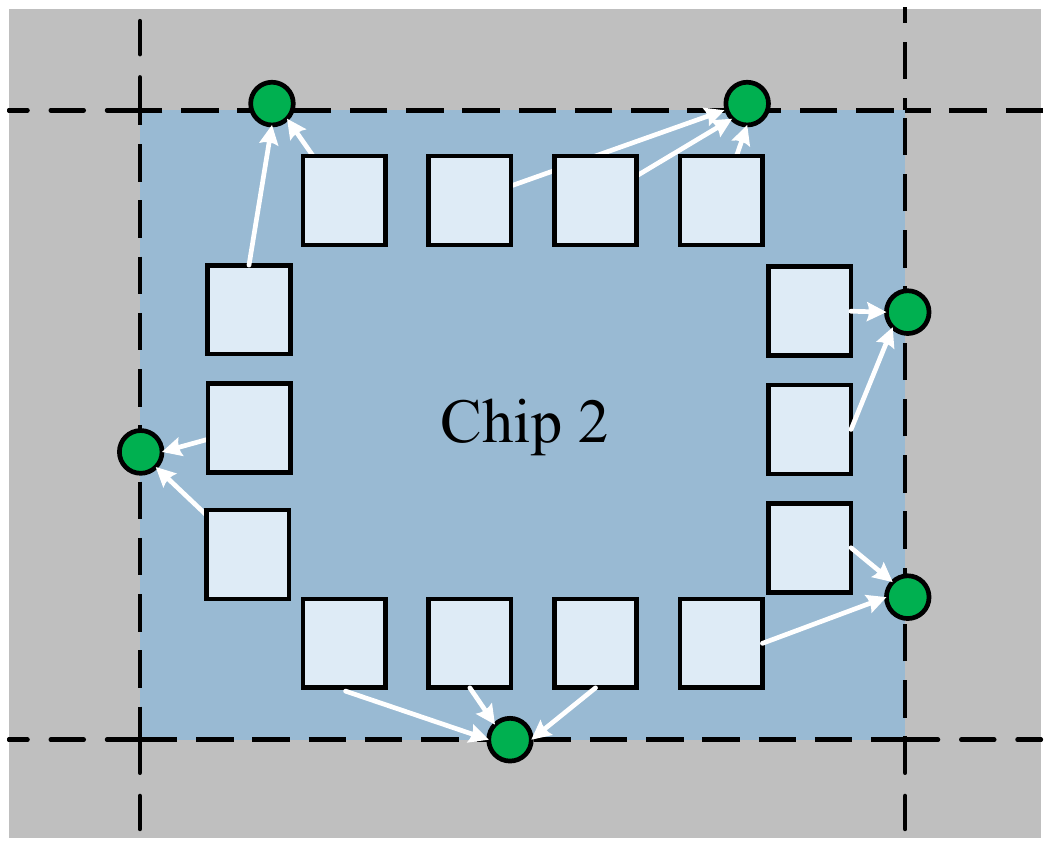}
		\caption{}
		\label{f:routingregion_c}
	\end{subfigure}
	
	\caption{(a) The layout is partitioned into irregular grids using the method in~\cite{ohtsuki1985gridless}. (b) Rectangular grids are merged following Ho et al.~\cite{lee2012obstacle}, based on which the routing graph is constructed and routing demand is estimated using the Top-$K$ path algorithm. (c) Pins inside fan-in regions are mapped to their nearest routing-graph vertices.}
	\label{f:routingregion}
\end{figure*}

To better estimate the currently available routing resources, we first partition the routing region into rectangular grids using the method in~\cite{ohtsuki1985gridless}. As shown in Figure~\ref{f:routingregion}(\subref{f:routingregion_a}), for each fan-in region, four extension lines are generated from its corner points along the four orthogonal directions, i.e., $0^\circ$, $90^\circ$, $180^\circ$, and $270^\circ$. Each extension line stops when it reaches another fan-in region or the boundary of the routing plane. In this way, the entire routing region is divided into a set of rectangular grids.

After the initial partition, to avoid excessive grid fragmentation, we adopt the method proposed by Ho et al.~\cite{lee2012obstacle} to merge rectangular grids and then further construct a routing graph. The routing graph is denoted as $G(V,E)$, where $V$ contains the midpoints of the common boundaries shared by adjacent routing grids. For two vertices $v_i$ and $v_j$, if the line segment connecting them does not overlap with any grid boundary, an edge $e_{ij}=(v_i,v_j)$ is added to $E$. Figure~\ref{f:routingregion}(\subref{f:routingregion_b}) shows the construction result of the routing graph.

Before path search, each pin is mapped to its nearest graph vertex. As shown in Figure~\ref{f:routingregion}(\subref{f:routingregion_c}), several pins may be located inside a fan-in region. During routing graph construction, each pin is mapped to the nearest vertex according to the Euclidean distance between the pin and graph vertices, and a pin mapping table is then constructed. In this way, the original pin-to-pin path search problem is transformed into a vertex-to-vertex path search problem on the routing graph. It should be noted that multiple pins may be mapped to the same vertex. This means that one path search result can be reused for multiple pin selections, thereby reducing the computational cost of congestion estimation. Meanwhile, each pin is only mapped to vertices on the boundary of its own fan-in region. Therefore, during the actual search process, path search only needs to be performed between boundary vertices of fan-in regions. Then, based on Yen's Top-$K$ shortest path algorithm~\cite{yen1971finding}, we generate $K$ candidate paths between pairs of fan-in boundary vertices. Figure~\ref{f:routingregion}(\subref{f:routingregion_b}) shows an example with $K=4$, where the four generated candidate paths are highlighted in red.
After path search, the algorithm counts the weighted number of paths passing through each vertex, denoted as $\mathrm{dem}(v)$. To avoid underestimating the available routing resources and blindly enlarging the inter-chip spacing, we assign different weights to paths with different ranks. Specifically, for the $K$ candidate paths of a net, paths generated earlier are regarded as higher-ranked candidates because they have shorter original or penalized path costs. The routing demand of vertex $v$ is then calculated as

\begin{equation}
	\mathrm{dem}(v)
	=
	\sum_{n\in N}
	\sum_{k=1}^{K}
	w_k \cdot \mathbb{I}\left(v \in P_n^{(k)}\right),
	\label{eq:vertex_demand}
\end{equation}
where $N$ denotes the set of signal nets, $P_n^{(k)}$ denotes the $k$-th diversity-aware candidate path of net $n$, and $w_k$ is the normalized weight assigned to that path and the weights satisfy $w_1 \ge w_2 \ge \cdots>w_K$ and $\sum_{k=1}^{K}w_k=1$. The indicator function $\mathbb{I}\left(v \in P_n^{(k)}\right)$ indicates whether vertex $v$ is included in the $k$-th candidate path of net $n$, and is defined as
\begin{equation}
	\mathbb{I}\left(v \in P_n^{(k)}\right)
	=
	\begin{cases}
		1, & v \in P_n^{(k)}, \\
		0, & v \notin P_n^{(k)} .
	\end{cases}
	\label{eq:indicator_function}
\end{equation}

To estimate the routing capacity of each vertex, we define the routing pitch $\rho$ as the minimum center-to-center distance between two adjacent wires. It is determined by the required wire spacing $s$ and the wire width $s_w$, i.e.,
\begin{equation}
	\rho = s + s_w ,
	\label{eq:routing_pitch}
\end{equation}
where $s$ and $s_w$ denote the wire spacing and wire width required by the design rules, respectively. Then, the routing capacity of vertex $v$ is defined as
\begin{equation}
	\mathrm{cap}(v)=\frac{l_v}{\rho},
	\label{eq:vertex_capacity}
\end{equation}
where $l_v$ denotes the length of the common boundary represented by vertex $v$.

Finally, the congestion of vertex $v$ is calculated as
\begin{equation}
	\mathrm{Congestion}(v)
	=
	\max\left(0,\mathrm{dem}(v)-\mathrm{cap}(v)\right).
	\label{eq:vertex_congestion}
\end{equation}

By defining the congestion of vertices in the routing graph, we can clearly characterize the current congestion distribution of the routing resources.

\subsubsection{Gradient-Based Routability Optimization:}

In this stage, we further construct a congestion propagation path. We map routing congestion to the available routing channels determined by the relative positions of chips and propagate congestion information back to chip center locations through differentiable loss functions. In this way, the optimizer can guide chip movement to release routing resources in critical regions.

Specifically, we first define chip boundaries according to the current chip locations and sizes. For chip $c_i$, its left, right, bottom, and top boundaries are respectively defined as
\begin{equation}
	\left\{
	\begin{aligned}
		L_i(\mathbf{c}) &= x_i - \frac{w_i}{2}, \\
		R_i(\mathbf{c}) &= x_i + \frac{w_i}{2}, \\
		B_i(\mathbf{c}) &= y_i - \frac{h_i}{2}, \\
		T_i(\mathbf{c}) &= y_i + \frac{h_i}{2}.
	\end{aligned}
	\right.
	\label{eq:chip_boundary}
\end{equation}
where $(x_i,y_i)$ denotes the center location of chip $c_i$, and $w_i$ and $h_i$ denote its width and height, respectively.

Given the chip boundaries, when a routing-demand vertex $v$ lies in a
horizontal or vertical routing channel between two chips, the length of the
associated routing channel is determined by the distance between the
corresponding adjacent chip boundaries. Let $o_v \in \{\mathrm{H}, \mathrm{V}\}$
denote the orientation of the routing channel containing $v$, with
$o_v=\mathrm{H}$ and $o_v=\mathrm{V}$ indicating horizontal and vertical
routing channels, respectively. The routing-channel length associated with
vertex $v$ is defined as
\begin{equation}
	l_v =
	\begin{cases}
		x_v^{+}-x_v^{-}, & o_v=\mathrm{H}, \\
		y_v^{+}-y_v^{-}, & o_v=\mathrm{V},
	\end{cases}
	\label{eq:guide_segment_length}
\end{equation}
where $x_v^{-}$ and $x_v^{+}$, or $y_v^{-}$ and $y_v^{+}$, denote the boundary coordinates of the two ends of the routing channel. A shorter $l_v$ indicates a narrower local routing channel. 

To impose constraints only on regions with actual routing pressure, we define the set of routing vertices with positive routing demand as
\begin{equation}
	V^{+}=\{v\in V \mid \mathrm{dem}(v)>0\},
	\label{eq:positive_demand_vertices}
\end{equation}

For these regions, we construct two loss terms according to the congestion severity, which are used to handle severe blockage and regular congestion, respectively. The first term is the zero-capacity loss, which penalizes the case where routing demand exists but the corresponding routing channel is too short:
\begin{equation}
	A_0 =
	\sum_{v\in V^{+}}
	\left[
	\max\left(0,\frac{\rho-l_v}{\rho}\right)
	\right]^2 ,
	\label{eq:zero_capacity_loss}
\end{equation}
When $l_v \ge \rho$, vertex $v$ has at least a basic routing space, and this loss becomes zero. When $l_v < \rho$, the loss increases as the routing channel length decreases. This term mainly prevents critical routing channels from being compressed to nearly zero width during optimization, thus preserving basic feasible routing space for all demand regions.

The second term is the over-capacity loss, which describes regular congestion where the routing channel length is insufficient to accommodate the corresponding routing demand:
\begin{equation}
	A_1 =
	\sum_{v\in V^{+}}
	\left[
	\max\left(0,\frac{\mathrm{dem}(v)\rho-l_v}{\rho}\right)
	\right]^2 .
	\label{eq:over_capacity_loss}
\end{equation}
Different from $A_0$, $A_1$ explicitly considers the actual routing demand $\mathrm{dem}(v)$. When the routing demand of a region is high, the required routing channel length should be increased accordingly. When the routing demand is low, excessive spacing does not need to be reserved. Therefore, $A_1$ can adaptively adjust the required channel width according to local routing demand, guiding the optimizer to prioritize high-demand regions and alleviate local routing bottlenecks.

Finally, the total loss of this stage is defined as
\begin{equation}
	\mathcal{L}_2 = \alpha_0 A_0 + \alpha_1 A_1 ,
	\label{eq:congestion_propagation_loss}
\end{equation}
where $\alpha_0$ and $\alpha_1$ are the weights of the zero-capacity loss and the over-capacity loss, respectively. The former emphasizes the preservation of basic routing space to avoid complete blockage of critical channels, while the latter emphasizes the matching between routing resources and routing demands, so that high-demand regions can obtain larger available routing channels.

To improve the stability of the optimization process, we further adopt a two-stage training strategy. In the first stage, only $A_0$ is optimized, ensuring that all vertices with routing demand obtain at least a basic routing channel length. This avoids unstable gradient directions caused by severe blockage at the beginning of optimization. In the second stage, $A_1$ is further optimized based on the first-stage result. The chip locations are continuously adjusted according to the routing demands of different regions, so that high-demand regions can obtain sufficient routing space. Through this coarse-to-fine optimization strategy, the congestion loss can be back-propagated along the differentiable path from routing channel length to chip boundaries and then to chip center coordinates. As a result, the optimizer can adjust chip locations, thereby enlarging the available space of critical routing channels and improving the overall RDL routability.

\section{Experimental Results}
\subsection{Experimental Setup}
The proposed package floorplanning algorithm is implemented using C++/CUDA and Python. Specifically, the crossing-aware pin assignment module is implemented in C++/CUDA, while the differentiable wirelength minimization and differentiable routability maximization modules are implemented in Python. All experiments are conducted on a workstation equipped with 16 GB of memory, an AMD Ryzen 9 7945HX CPU running at 2.50 GHz, and an NVIDIA GeForce RTX 4060 GPU.
We evaluate the proposed method on ten benchmark circuits which were originally introduced in~\cite{wen2022via} and have been widely adopted in subsequent RDL routing studies, such as~\cite{li2026redistribution,xu2025hierarchical}.  Derived from practical FOWLP applications, these benchmarks exhibit high density and severe congestion, making their complexity comparable to that of real designs. The detailed benchmark information is reported in Table~\ref{t:1}, where Area, $|C|$, $|L|$, $|I/O|$, and $|N|$ denote the routing area per layer, the number of chips, the number of routing layers, the number of I/O pads, and the number of inter-chip nets, respectively.

\begin{table*}[h]
	\centering
	\caption{Benchmark statistics.}
	\label{t:1}
	\setlength{\tabcolsep}{5pt}
	\renewcommand{\arraystretch}{1.05}
	\begin{tabular}{|c|c|c|c|c|c|}
		\hline
		Circuits & Area($\mu$m$ \times \mu$m) & $|C|$ & $|L|$ & $|I/O|$ & $|N|$ \\
		\hline
		\noalign{\vskip 2pt}
		\hline
		dense1 & $5000{\times}5000$ & 2 & 2 & 44 & 22 \\
		\hline
		dense2 & $7500{\times}7500$ & 3 & 3 & 92 & 46 \\
		\hline
		dense3 & $6000{\times}4000$ & 5 & 3 & 164 & 80 \\
		\hline
		dense4 & $10000{\times}5000$ & 6 & 3 & 222 & 111 \\
		\hline
		dense5 & $10000{\times}10000$ & 9 & 4 & 522 & 261 \\
		\hline
		pkg1 & $5000{\times}5000$ & 3 & 2 & 20 & 10 \\
		\hline
		pkg2 & $6000{\times}6000$ & 4 & 2 & 56 & 28 \\
		\hline
		pkg3 & $8000{\times}8000$ & 7 & 2 & 108 & 54 \\
		\hline
		pkg4 & $12000{\times}6000$ & 12 & 2 & 180 & 90 \\
		\hline
		pkg5 & $16000{\times}12000$ & 20 & 2 & 492 & 246 \\
		\hline
	\end{tabular}
\end{table*}

\subsection{Experimental Results and Analysis}
To evaluate the overall performance of the proposed method, we use the state-of-the-art RDL router~\cite{li2026redistribution} to validate the floorplanning results. For comparison, we reproduce the algorithm proposed in~\cite{lin2023routability}. To the best of our knowledge, this algorithm is the current state-of-the-art routability-driven package floorplanning algorithm. Based on RUDY~\cite{spindler2007routing}, \cite{lin2023routability} proposes an X-architecture-oriented congestion model, which accounts for the occupation of diagonal routing grids by $135^\circ$ nets and thus provides a more accurate estimation of routing congestion in SiP designs. However, \cite{lin2023routability} does not consider pin assignment. To demonstrate the benefit of jointly considering pin assignment during floorplanning, we further construct a baseline denoted as \cite{lin2023routability}+PA, where the proposed pin assignment algorithm is incorporated into the framework of~\cite{lin2023routability}. Table~\ref{t:2} compares the proposed method with~\cite{lin2023routability} in terms of HPWL and runtime. Table~\ref{t:3} further reports the routability and routed wirelength after applying the RDL router in~\cite{li2026redistribution}.

As shown by the normalized HPWL values in Table~\ref{t:2}, \cite{lin2023routability}+PA has a normalized HPWL Comp. value of 0.76, which is 24\% lower than that of the proposed method. This is because the proposed method reserves sufficient fan-out routing channels by increasing the spacing between chips, which in turn ensures complete routability in the subsequent routing stage. In addition, \cite{lin2023routability}+PA reduces the normalized HPWL Comp. value from 1.57 to 0.76 compared with~\cite{lin2023routability}, demonstrating the effectiveness of the proposed pin assignment algorithm. Given a fixed floorplanning result, the proposed pin assignment algorithm can significantly reduce HPWL by adjusting the pin locations at both ends of each net. 

In terms of normalized runtime, compared with~\cite{lin2023routability}, adding the proposed pin-assignment stage increases the Comp. value from 0.57 to 0.65, i.e., an absolute increase of only 0.08 in the normalized runtime metric, indicating that the proposed pin assignment algorithm introduces limited computational overhead. Compared with \cite{lin2023routability}+PA, the Comp. value of the proposed method further increases to 1.00, with an absolute increase of 0.35. This is mainly due to the accurate congestion estimation method designed for the fan-out routing style. When a chip moves to the boundary, the routing graph needs to be repartitioned and the related routing paths need to be re-estimated, which introduces additional runtime overhead.
\begin{table}[h]
	\centering
	\small
	\caption{Comparison with work in ~\cite{lin2023routability} in HPWL and runtime.}
	\label{t:2}
	\setlength{\tabcolsep}{3pt}
	\renewcommand{\arraystretch}{1.05}
	\begin{tabular}{|*{7}{Z|}}
		\hline
		\multirow{2}{*}{Circuits} 
		& \multicolumn{3}{c|}{HPWL ($\mu$m)} 
		& \multicolumn{3}{c|}{Runtime (sec.)} \\
		\cline{2-7}
		& Ours & Lin~\cite{lin2023routability} & Lin~\cite{lin2023routability}+PA & Ours & Lin~\cite{lin2023routability} & Lin~\cite{lin2023routability}+PA \\
		\hline
		\noalign{\vskip 2pt}
		\hline
		dense1 & 7647.31 & 66762.70 & 10060.01 & 7.14 & 17.08 & 17.97 \\
		\hline
		dense2 & 34063.72 & 227527.04 & 35174.13 & 7.45 & 13.77 & 14.53 \\
		\hline
		dense3 & 117152.30 & 185895.96 & 113694.30 & 24.18 & 15.00 & 15.52 \\
		\hline
		dense4 & 266836.30 & 311811.05 & 172586.81 & 23.78 & 15.39 & 17.91 \\
		\hline
		dense5 & 1060787.56 & 1232124.39 & 761853.21 & 35.26 & 18.42 & 24.29 \\
		\hline
		pkg1 & 1702.24 & 11179.36 & 1658.51 & 16.09 & 14.03 & 14.09 \\
		\hline
		pkg2 & 17364.07 & 47128.27 & 5667.16 & 25.16 & 14.02 & 14.63 \\
		\hline
		pkg3 & 27327.70 & 83373.90 & 19729.12 & 18.61 & 14.84 & 15.80 \\
		\hline
		pkg4 & 44204.90 & 124631.76 & 41842.19 & 72.87 & 15.77 & 21.33 \\
		\hline
		pkg5 & 245396.31 & 576273.32 & 227845.02 & 46.75 & 18.78 & 24.22 \\
		\hline
		\noalign{\vskip 2pt}
		\hline
		Comp. & 1.00 & 1.57 & 0.76 & 1.00 & 0.57 & 0.65 \\
		\hline
	\end{tabular}
\end{table}

As shown in Table~\ref{t:3}, the proposed method achieves 100\% routability on all test cases, whereas ~\cite{lin2023routability} achieves full routability only on pkg1, and \cite{lin2023routability}+PA achieves full routability only on dense1, dense2, and pkg1. The main reason is that the congestion model in~\cite{lin2023routability} treats all partitioned routing grids equally, making it difficult to accurately capture the actual congestion of routing channels under the fan-out routing style. Meanwhile, compared with~\cite{lin2023routability}, \cite{lin2023routability}+PA shows better routability, indicating that the proposed pin assignment algorithm can select closer pins for signal nets, thereby preventing nets from crossing multiple fan-in regions and reducing the risk of routing conflicts.

Finally, we report the routed wirelength obtained after detailed routing. Note that the values in Table~\ref{t:3} are calculated only for successfully routed nets, and the symbol > denotes a lower bound. For pkg1, where all algorithms achieve full routability, the proposed method still obtains the shortest routed wirelength, reducing it by approximately 84\% compared with Lin~\cite{lin2023routability} and by approximately 11\% compared with Lin~\cite{lin2023routability}+PA. For dense1 and dense2, where Lin~\cite{lin2023routability}+PA also achieves full routability, the proposed method reduces the routed wirelength by 23\% and 5\%, respectively. These results demonstrate that the proposed congestion estimation model can accurately estimate the required capacity of routing channels and reduce wirelength while ensuring complete routability.
\begin{table}[h]
	\centering
	\small
	\caption{Comparison with work in ~\cite{lin2023routability} in routability and wirelength.}
	\label{t:3}
	\setlength{\tabcolsep}{3pt}
	\renewcommand{\arraystretch}{1.05}
	\begin{tabular}{|*{7}{Z|}}
		\hline
		\multirow{2}{*}{Circuits} 
		& \multicolumn{3}{c|}{Routability} 
		& \multicolumn{3}{c|}{Wirelength ($\mu$m)} \\
		\cline{2-7}
		& Ours & Lin~\cite{lin2023routability} & Lin~\cite{lin2023routability}+PA & Ours & Lin~\cite{lin2023routability} & Lin~\cite{lin2023routability}+PA \\
		\hline
		\noalign{\vskip 2pt}
		\hline
		dense1 & 1.00 & 0.77 & 1.00 & 7837.64 & $>$59776.80 & 10211.25 \\
		\hline
		dense2 & 1.00 & 0.87 & 1.00 & 35604.24 & $>$306291.37 & 37745.55 \\
		\hline
		dense3 & 1.00 & 0.46 & 0.58 & 149950.34 & $>$144267.92 & $>$76002.20 \\
		\hline
		dense4 & 1.00 & 0.46 & 0.73 & 265920.84 & $>$235069.61 & $>$162151.45 \\
		\hline
		dense5 & 1.00 & 0.62 & 0.87 & 1167257.58 & $>$1284104.67 & $>$1044855.04 \\
		\hline
		pkg1 & 1.00 & 1.00 & 1.00 & 2328.74 & 15145.92 & 2634.28 \\
		\hline
		pkg2 & 1.00 & 0.54 & 0.93 & 15276.87 & $>$43637.20 & $>$4776.77 \\
		\hline
		pkg3 & 1.00 & 0.65 & 0.98 & 32884.05 & $>$120744.50 & $>$38233.16 \\
		\hline
		pkg4 & 1.00 & 0.61 & 0.91 & 59082.14 & $>$127958.80 & $>$60030.62 \\
		\hline
		pkg5 & 1.00 & 0.54 & 0.88 & 211486.06 & $>$493355.78 & $>$208579.34 \\
		\hline
		\noalign{\vskip 2pt}
		\hline
		Comp. & 1.00 & 0.65 & 0.89 & 1.00 & - & - \\
		\hline
	\end{tabular}
\end{table}

To evaluate the performance of the proposed pin assignment algorithm, we implement the pin assignment algorithms used in~\cite{li2016pin} and~\cite{zhuang2026adaptive}. In particular, the MIS computation is performed using KaMIS~\cite{DBLP:journals/heuristics/LammSSSW17}. The method in~\cite{li2016pin} adopts a simulated annealing (SA)-based pin assignment algorithm, which significantly reduces runtime by constructing a high-quality initial solution and partitioning large-scale pin arrays into several subregions during the search process. The method in~\cite{zhuang2026adaptive} proposes a maximum independent set (MIS)-based pin assignment algorithm. It first generates multiple candidate pin assignment solutions for each net and then selects a set of non-conflicting assignments through MIS. If some nets remain unassigned, this process is iteratively repeated until a complete solution is obtained.

\begin{table}[h]
	\centering
	\small
	\caption{Comparison with SA-based~\cite{li2016pin} and MIS-based~\cite{zhuang2026adaptive} pin assignment methods in HPWL and runtime.}
	\label{tab:pa_comparison}
	\setlength{\tabcolsep}{3pt}
	\renewcommand{\arraystretch}{1.05}
	\begin{tabular}{|*{7}{Z|}}
		\hline
		\multirow{2}{*}{Circuits}
		& \multicolumn{3}{c|}{HPWL ($\mu$m)}
		& \multicolumn{3}{c|}{Runtime (sec.)} \\
		\cline{2-7}
		& Ours & SA~\cite{li2016pin} & MIS~\cite{zhuang2026adaptive}
		& Ours & SA~\cite{li2016pin} & MIS~\cite{zhuang2026adaptive} \\
		\hline
		\noalign{\vskip 2pt}
		\hline
		dense1 & 7647.31 & 7661.85 & 7649.25 & 7.14 & 7.50 & 18.63 \\
		\hline
		dense2 & 34063.72 & 102576.74 & 52638.54 & 7.45 & 8.85 & 20.89 \\
		\hline
		dense3 & 117152.30 & 184200.77 & 159322.13 & 24.18 & 26.50 & 35.13 \\
		\hline
		dense4 & 266836.30 & 402532.14 & 286393.01 & 23.78 & 27.82 & 35.91 \\
		\hline
		dense5 & 1060787.56 & 1138715.10 & 1087414.45 & 35.26 & 56.62 & 50.17 \\
		\hline
		pkg1 & 1702.24 & 4185.63 & 1832.04 & 16.09 & 16.16 & 23.57 \\
		\hline
		pkg2 & 17364.07 & 22091.34 & 17002.41 & 25.16 & 25.47 & 35.14 \\
		\hline
		pkg3 & 27327.70 & 40455.14 & 44098.04 & 18.61 & 19.21 & 28.62 \\
		\hline
		pkg4 & 44204.90 & 85567.41 & 64063.57 & 72.87 & 74.47 & 82.39 \\
		\hline
		pkg5 & 245396.31 & 221408.37 & 179984.32 & 46.75 & 59.62 & 57.81 \\
		\hline
		\noalign{\vskip 2pt}
		\hline
		Comp. & 1.00 & 1.21 & 1.04 & 1.00 & 1.16 & 1.40 \\
		\hline
	\end{tabular}
\end{table}

The experimental results are shown in Table~\ref{tab:pa_comparison}. In terms of runtime, The proposed method achieves the shortest runtime on all benchmarks. Its advantage is relatively modest on small-scale cases but becomes more pronounced on larger designs. For small-scale cases, the runtime overhead of the proposed method mainly comes from the inherent cost of the differentiable optimization framework, such as GPU kernel launches and frequent CPU--GPU data transfer during forward and backward propagation. In contrast, the SA-based algorithm in~\cite{li2016pin} can quickly generate feasible solutions with low overhead, and therefore achieves shorter runtime on some small-scale test cases. However, as the circuit scale increases, the computation is gradually dominated by large-scale parallelizable operations, allowing the GPU to better amortize the fixed overhead and fully exploit its parallel computing capability. As a result, the proposed method shows a more significant runtime advantage on larger designs. For example, compared with the result of~\cite{li2016pin} on dense5, the proposed method reduces the runtime from 56.62 s to 35.26 s, corresponding to a reduction of approximately 37.72\% and a speedup of about 1.61$\times$. This demonstrates the good scalability of the proposed method for large-scale multi-chip and multi-layer RDL designs.

In terms of HPWL, SA produces 21\% higher HPWL than the proposed method. This improvement is mainly because the proposed algorithm maintains particle diversity during the particle update process by dividing particles into multiple subgroups and particle types, thereby avoiding premature convergence to local optima. In addition, the MIS-based method achieves comparable or even better results than the proposed method on dense1, pkg1 and pkg2. This is mainly because pin competition is relatively weak in these test cases, and conflicts among the optimal pin choices of different nets are limited. In each iteration, the MIS-based method attempts to select the best pin assignment for all unassigned nets. However, the optimal pin choices of different nets may conflict with each other. In this case, the algorithm completes the pin assignment for a subset of nets according to their priorities and leaves the remaining unassigned nets to the next iteration. Therefore, when conflicts among the optimal pin choices of different nets are rare, the MIS-based method can obtain high-quality solutions. Ideally, if the optimal pin choices of all nets are mutually non-conflicting, the method can directly obtain the optimal assignment.

Table~\ref{tab:pa_routability_wirelength} reports the results of an ablation study on the flightline-collision cost in the pin assignment stage. Both the proposed method and the variant without the flightline-collision cost achieve complete routability on all benchmarks, indicating that the final routability is mainly guaranteed by the overall floorplanning and routability optimization framework. However, the proposed method consistently obtains shorter routed wirelength on all test cases. Compared with the variant without FC, the routed wirelength is reduced by 4.66\% on average, with larger improvements on pkg2, pkg5, dense4, and pkg4. These results show that incorporating the flightline-collision cost does not merely serve as a routability filter, but helps generate pin assignments with fewer potential crossings, thereby reducing routing detours and improving the final routed wirelength.
\begin{table}[h]
	\centering
	\small
	\caption{Effectiveness of the flightline-collision cost in pin assignment.}
	\label{tab:pa_routability_wirelength}
	\setlength{\tabcolsep}{3pt}
	\renewcommand{\arraystretch}{1.05}
	\begin{tabular}{|*{6}{Z|}}
		\hline
		\multirow{2}{*}{Circuits}
		& \multicolumn{2}{c|}{Routability}
		& \multicolumn{3}{c|}{Wirelength ($\mu$m)} \\
		\cline{2-6}
		& w/ FC & \makecell{w/o FC}
		& w/ FC & \makecell{w/o FC} & Ratio (\%) \\
		\hline
		\noalign{\vskip 2pt}
		\hline
		dense1 & 1.00 & 1.00 & 7837.64 & 7845.17 & 0.10 \\
		\hline
		dense2 & 1.00 & 1.00 & 35604.24 & 35786.62 & 0.51 \\
		\hline
		dense3 & 1.00 & 1.00 & 149950.34 & 152321.99 & 1.56 \\
		\hline
		dense4 & 1.00 & 1.00 & 265920.84 & 285139.57 & 6.74 \\
		\hline
		dense5 & 1.00 & 1.00 & 1167257.58 & 1221731.14 & 4.46 \\
		\hline
		pkg1 & 1.00 & 1.00 & 2328.74 & 2377.48 & 2.05 \\
		\hline
		pkg2 & 1.00 & 1.00 & 15276.87 & 17853.30 & 14.43 \\
		\hline
		pkg3 & 1.00 & 1.00 & 32884.05 & 33613.63 & 2.17 \\
		\hline
		pkg4 & 1.00 & 1.00 & 59082.14 & 62692.43 & 5.76 \\
		\hline
		pkg5 & 1.00 & 1.00 & 211486.06 & 231865.96 & 8.79 \\
		\hline
		\noalign{\vskip 2pt}
		\hline
		Avg. & 1.00 & 1.00 & - & - & 4.66 \\
		\hline
	\end{tabular}
\end{table}

In the pin assignment stage, we employ GPU-based parallel computation to accelerate the pin assignment process. To evaluate the effectiveness of the GPU acceleration, we further compare the single-run runtime of the CPU and GPU versions of the proposed pin assignment algorithm. The experimental results are shown in Table~\ref{tab:cpu_gpu_runtime}. The proposed GPU implementation achieves significant speedup on all test cases. As the circuit scale increases, the runtime of the CPU version increases substantially, while the GPU version can better exploit its parallel computing capability. Across all test cases, the runtime of the GPU version remains within 3 seconds. On pkg5, the CPU version requires 470.28 s, whereas the GPU version takes only 2.85 s, achieving a speedup of 165.01$\times$. Over all test cases, the GPU version achieves an average speedup of 112.07$\times$. These results demonstrate that the proposed pin assignment algorithm has strong parallelism, and GPU-based parallel computation can effectively reduce the runtime, thereby meeting the requirement of fast optimization for large-scale circuits.

\begin{table}[h]
	\centering
\caption{Comparison of the single-run runtime of the CPU and GPU implementations for pin assignment.}
	\label{tab:cpu_gpu_runtime}
	\setlength{\tabcolsep}{10pt}
	\renewcommand{\arraystretch}{1.05}
	\begin{tabular}{|c|c|c|c|}
		\hline
		\multirow{2}{*}{Circuits}
		& \multicolumn{3}{c|}{Runtime (sec.)} \\
		\cline{2-4}
		& Ours (CPU) & Ours & Speedup \\
		\hline
		\noalign{\vskip 2pt}
		\hline
		dense1 & 50.97 & 0.70 & $72.81\times$ \\
		\hline
		dense2 & 86.34 & 1.18 & $73.17\times$ \\
		\hline
		dense3 & 133.86 & 1.47 & $91.06\times$ \\
		\hline
		dense4 & 159.21 & 1.36 & $117.07\times$ \\
		\hline
		dense5 & 270.96 & 2.82 & $96.09\times$ \\
		\hline
		pkg1 & 67.62 & 0.75 & $90.16\times$ \\
		\hline
		pkg2 & 99.28 & 0.92 & $107.91\times$ \\
		\hline
		pkg3 & 177.30 & 1.28 & $138.52\times$ \\
		\hline
		pkg4 & 322.65 & 1.91 & $168.93\times$ \\
		\hline
		pkg5 & 470.28 & 2.85 & $165.01\times$ \\
		\hline
		\noalign{\vskip 2pt}
		\hline
		Average & - & - & $112.07\times$ \\
		\hline
	\end{tabular}
\end{table}

The proposed method can generate better package floorplanning solutions for the following reasons.

\begin{itemize}
	\item Through the proposed pin assignment algorithm, the proposed method can select more appropriate pin combinations for signal nets. This not only reduces HPWL but also effectively decreases potential flightline collision, thereby reducing the actual wirelength in the subsequent routing stage.
	
	\item The proposed congestion estimation method can accurately estimate the actual utilization of routing resources under the fan-out routing style. Different from previous coarse-grained congestion estimation methods based on net density or bounding boxes, the proposed method directly models the occupation of routing channels according to the practical characteristics of fan-out routing. Therefore, it can more accurately capture local congestion in advanced packaging and provide effective guidance for subsequent gradient-based floorplanning optimization.
	
	\item The congestion in fan-out regions is estimated by the fan-out-aware congestion estimation model and then incorporated into a differentiable congestion loss for gradient-based optimization. In each iteration, the proposed method updates the congestion costs. By back-propagating the gradients through routing-channel lengths, the algorithm optimizes the widths of routing channels and gradually drives the layout toward lower congestion. As a result, the final floorplanning solution can provide sufficient routing resources to achieve complete routability.
\end{itemize}

\begin{figure}[!htbp]
	\centering
	
	\begin{subfigure}[b]{0.32\linewidth}
		\centering
		\includegraphics[width=\linewidth]{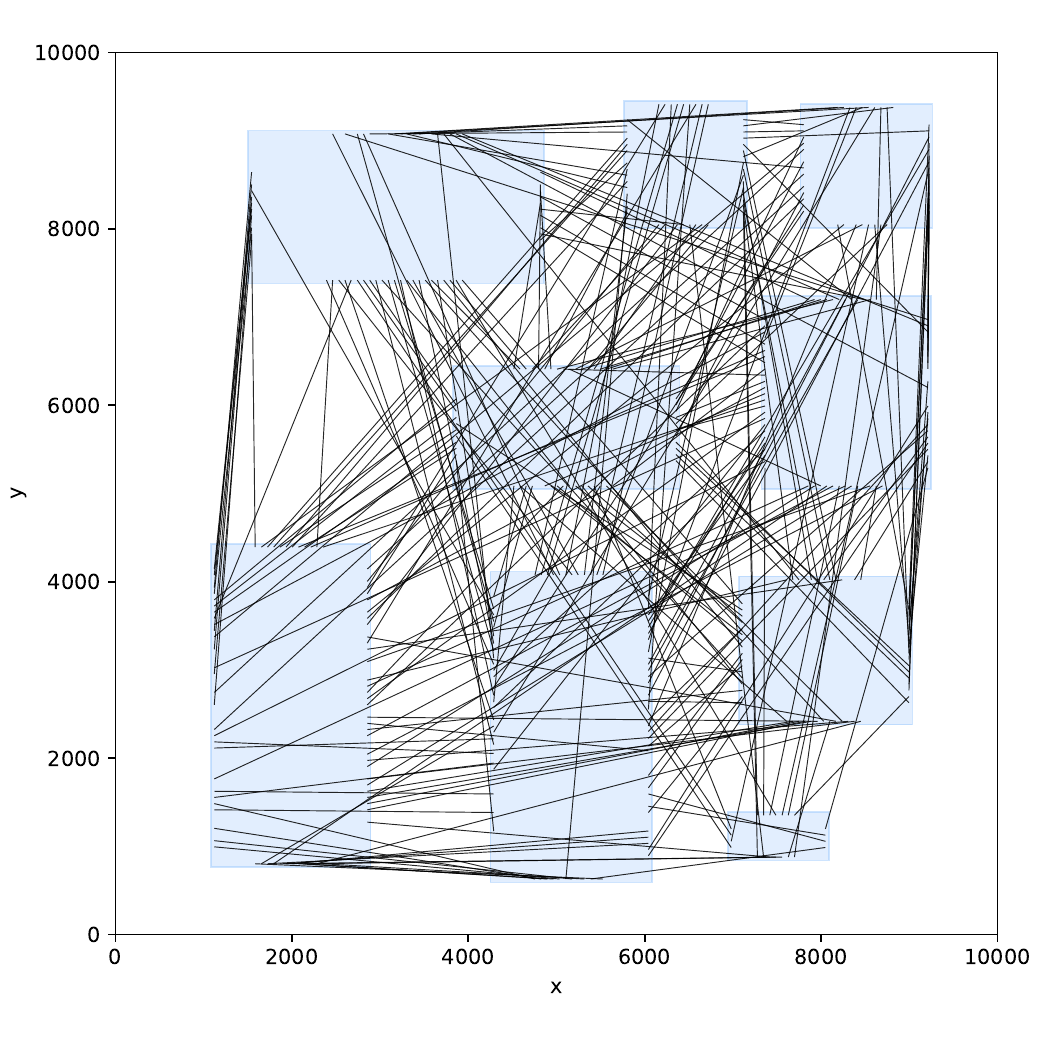}
		\caption{Initial Layout}
		\label{f:dense5_initial}
	\end{subfigure}
	\hfill
	\begin{subfigure}[b]{0.32\linewidth}
		\centering
		\includegraphics[width=\linewidth]{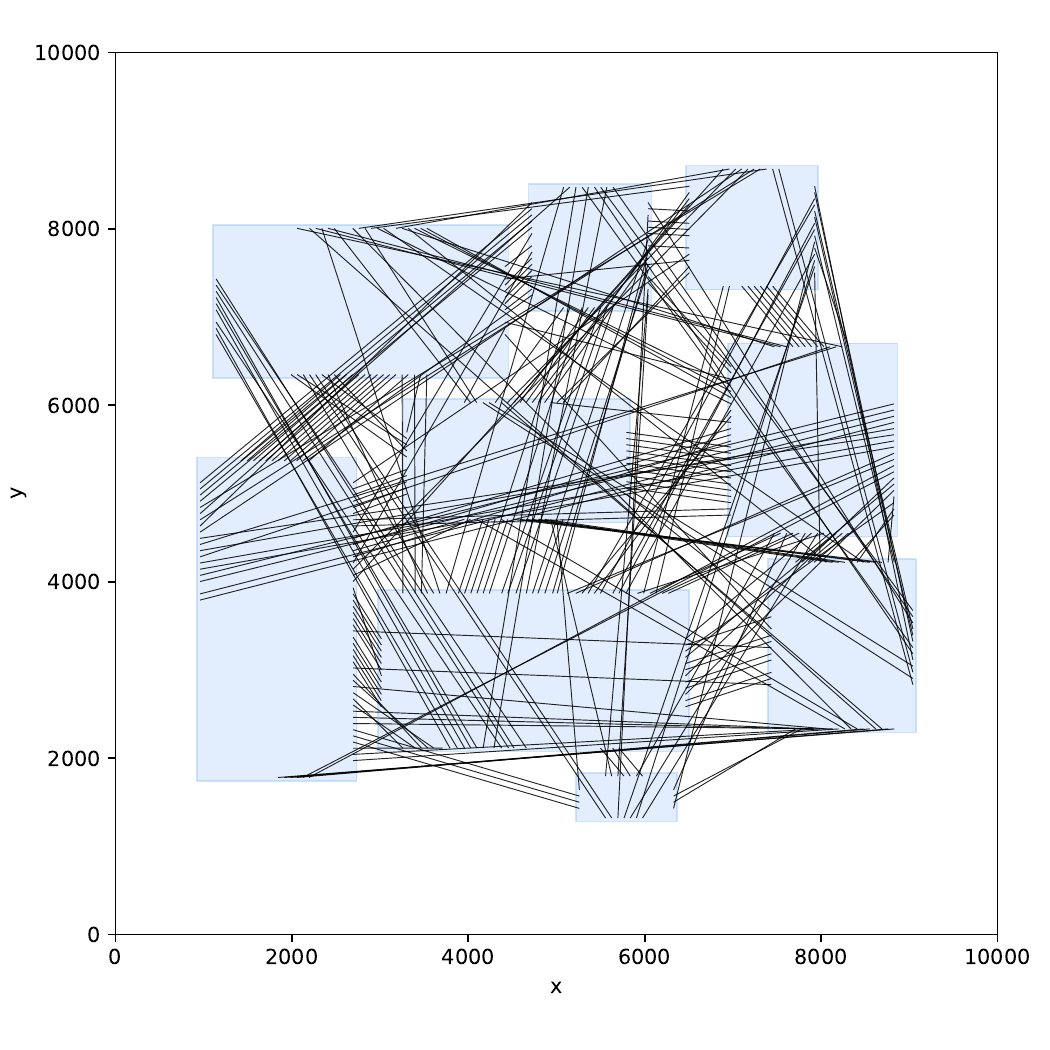}
		\caption{Floorplanning Result}
		\label{f:dense5_floorplanning}
	\end{subfigure}
	\hfill
	\begin{subfigure}[b]{0.32\linewidth}
		\centering
		\includegraphics[width=\linewidth]{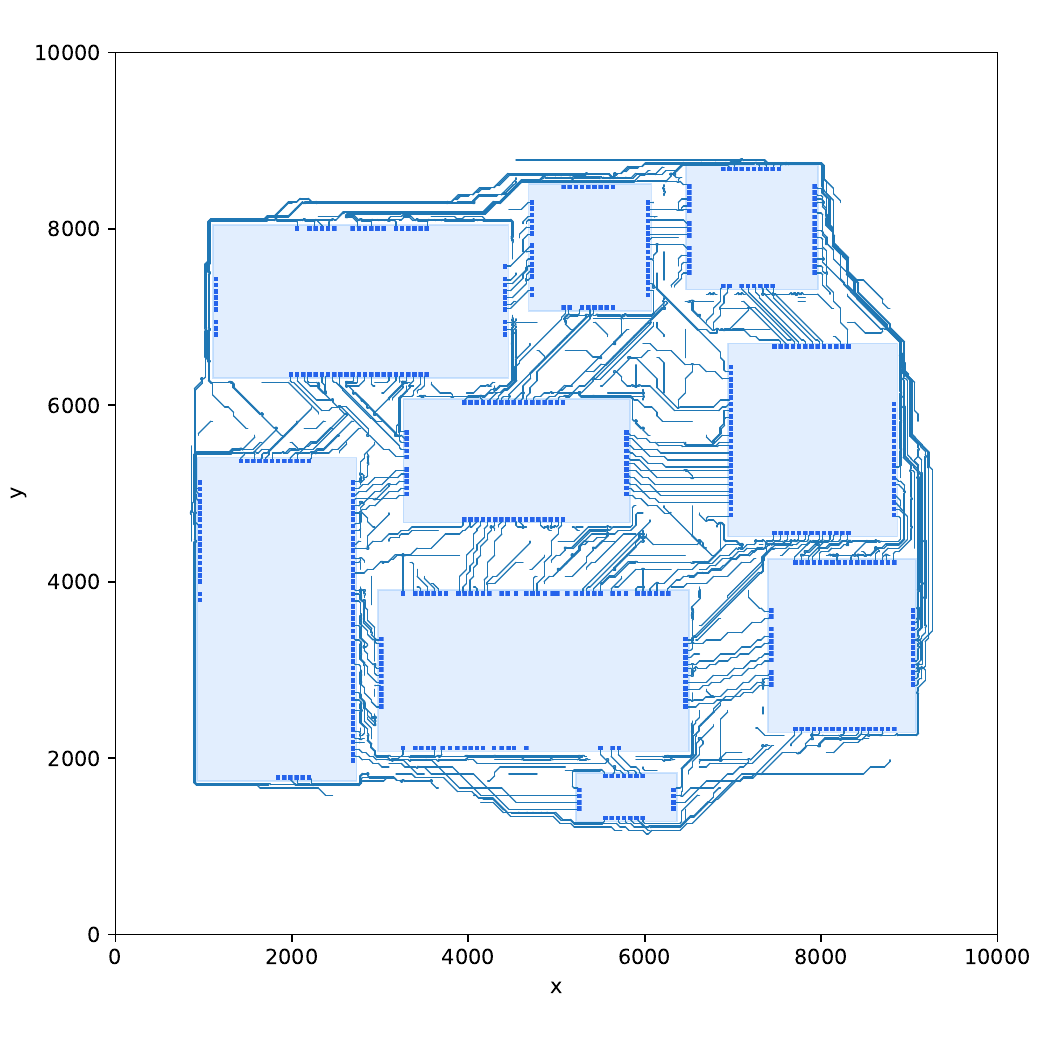}
		\caption{Layer 1}
		\label{f:dense5_layer1}
	\end{subfigure}
	
	\vspace{1mm}
	
	\begin{subfigure}[b]{0.32\linewidth}
		\centering
		\includegraphics[width=\linewidth]{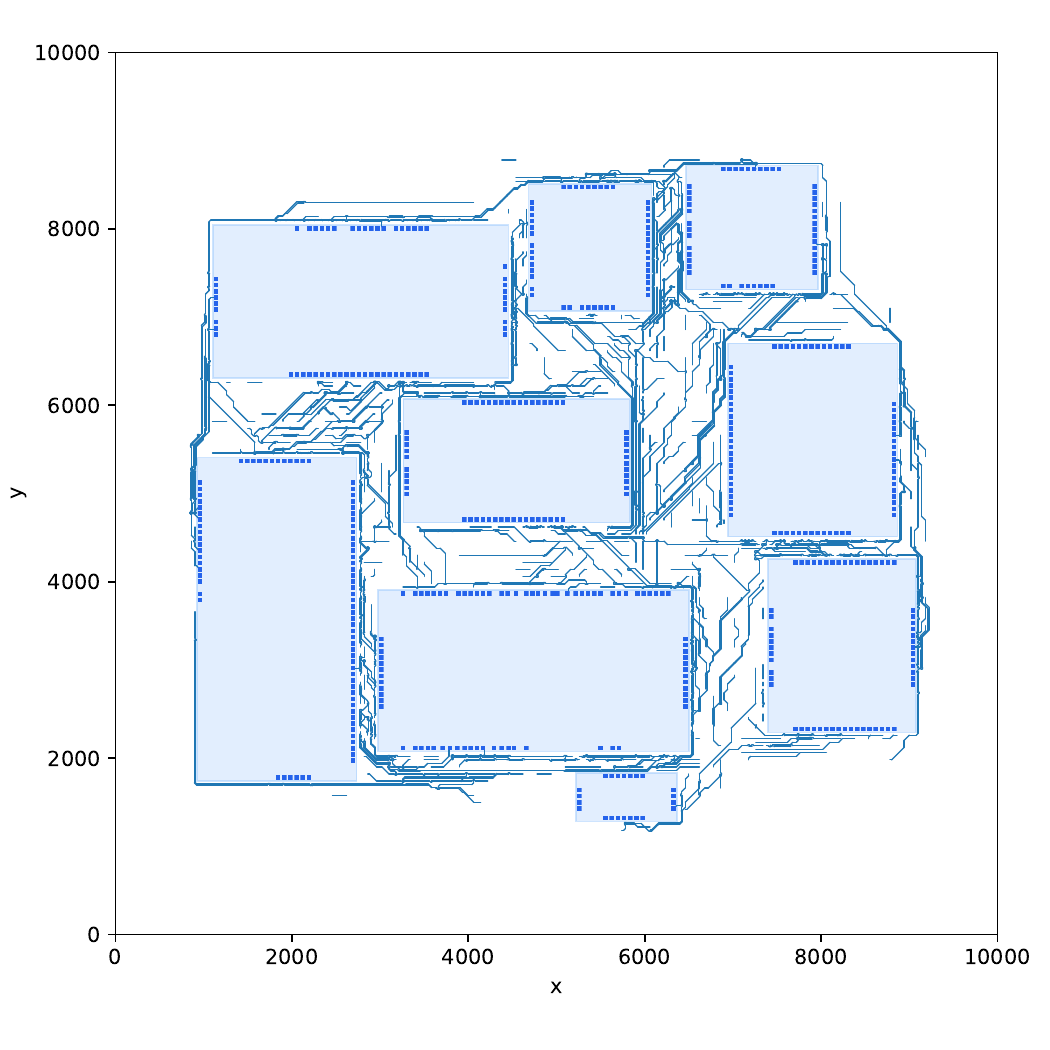}
		\caption{Layer 2}
		\label{f:dense5_layer2}
	\end{subfigure}
	\hfill
	\begin{subfigure}[b]{0.32\linewidth}
		\centering
		\includegraphics[width=\linewidth]{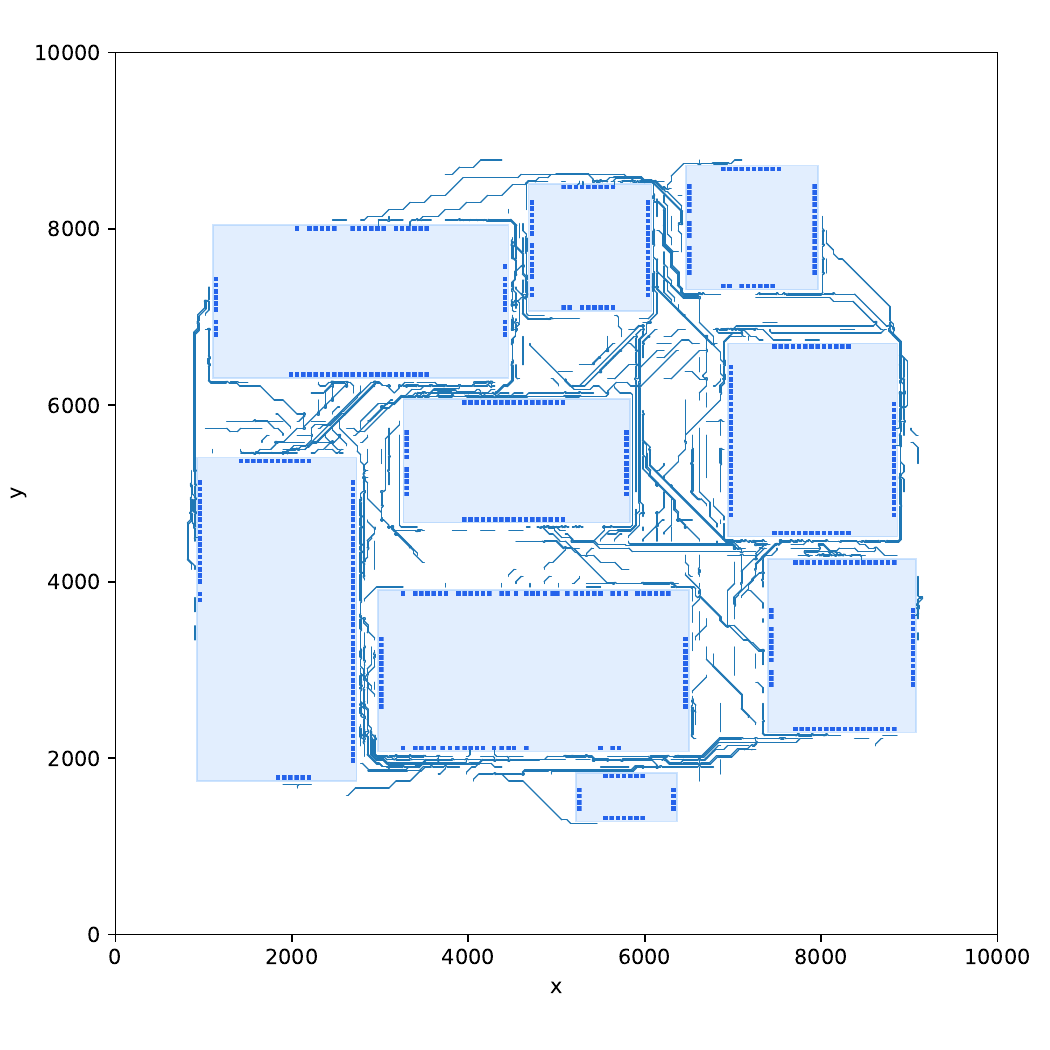}
		\caption{Layer 3}
		\label{f:dense5_layer3}
	\end{subfigure}
	\hfill
	\begin{subfigure}[b]{0.32\linewidth}
		\centering
		\includegraphics[width=\linewidth]{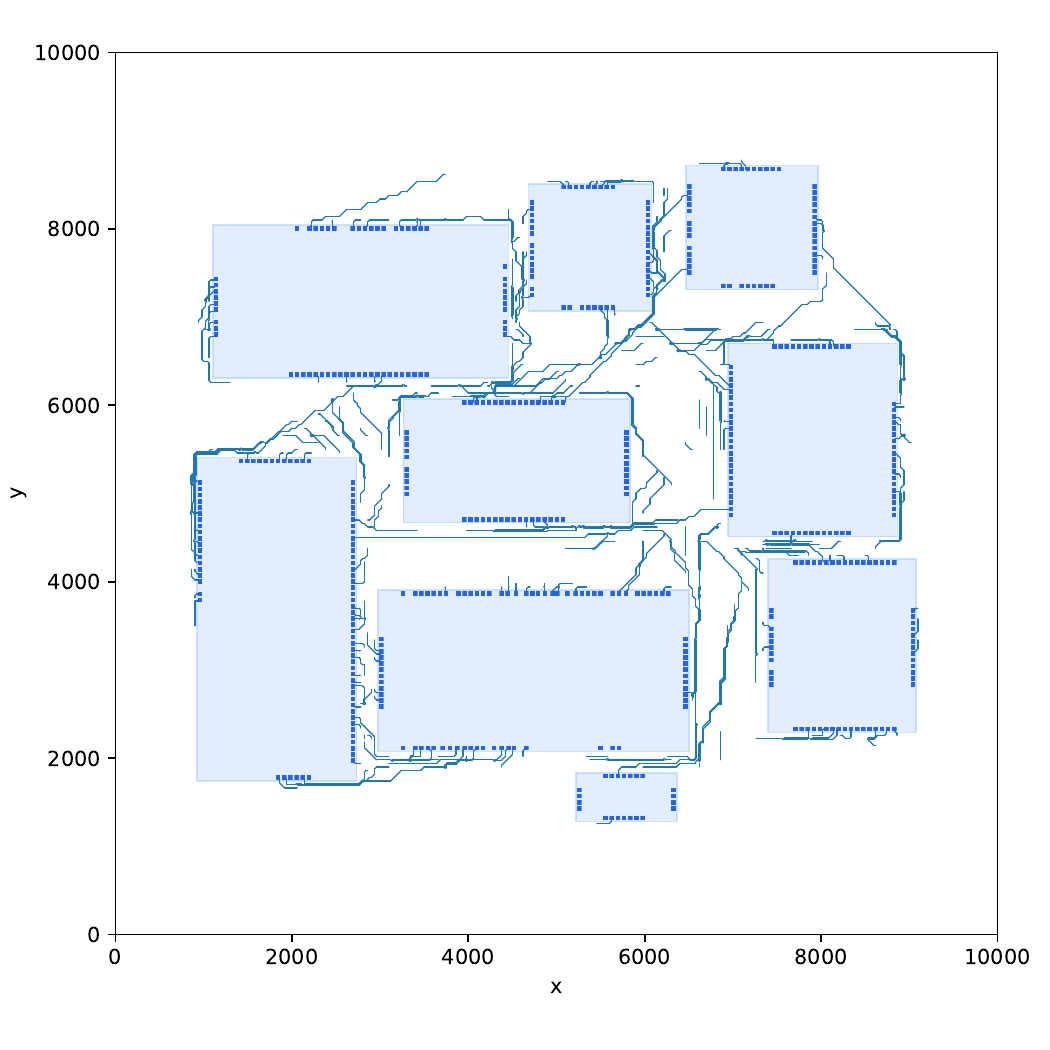}
		\caption{Layer 4}
		\label{f:dense5_layer4}
	\end{subfigure}
	
	\caption{Floorplanning and routing results of dense5.}
	\label{f:dense5_routing_results}
	
\end{figure}

Overall, different stages of the proposed method are seamlessly connected and efficiently handle different optimization tasks, enabling the routing stage to achieve 100\% routability with low routed wirelength. Figure~\ref{f:dense5_routing_results}(\subref{f:dense5_initial}) and Figure~\ref{f:dense5_routing_results}(\subref{f:dense5_floorplanning}) show the initial layout and the optimized floorplanning result of dense5, respectively, while Figure~\ref{f:dense5_routing_results}(\subref{f:dense5_layer1})--Figure~\ref{f:dense5_routing_results}(\subref{f:dense5_layer4}) show the final routing results. Dense5 is a highly congested benchmark case with nine chips, 522 I/O pads, and 261 inter-chip signal nets, which need to be connected within a $10{,}000~\mu\mathrm{m} \times 10{,}000~\mu\mathrm{m}$ routing region through multiple RDL layers. In addition, the initial layout of dense5 contains as many as 10,245 flightline collisions, making routing extremely difficult without optimization. After applying the proposed floorplanning algorithm, the number of flightline collision is reduced to 4,111, corresponding to a reduction of 59.87\%.

As shown in the routing results of Layers 1--4 in Figure~\ref{f:dense5_routing_results}(\subref{f:dense5_layer1})--Figure~\ref{f:dense5_routing_results}(\subref{f:dense5_layer4}), all routing channels are effectively utilized by the proposed method. Signal nets on each layer are successfully constructed without violating any design rules, demonstrating the effectiveness of the proposed method in improving routability for highly congested advanced packaging designs.

\section{Conclusion}
This paper presented a differentiable routability-driven package floorplanning with pin assignment algorithm for fan-out wafer-level packaging. To address the limited accuracy of conventional uniform-grid congestion models under the fan-out routing style, the proposed method jointly optimizes chip placement, chip orientation, pin assignment, and fan-out routing resources through a three-stage framework. First, a rotation-aware smooth HPWL formulation is used to optimize chip locations and legal orientations while preserving differentiability. Second, a crossing-aware pin assignment algorithm based on discrete particle swarm optimization is introduced to reduce both HPWL and potential flightline collision. Third, a fan-out-aware congestion estimation model is developed by partitioning available routing regions and estimating routing demand using Top-$K$ candidate paths, enabling congestion information to guide gradient-based floorplanning optimization. Experimental results show that the proposed method achieves 100\% routability on all benchmarks. Moreover, on fully routable cases, it reduces routed wirelength by a maximum of approximately 23\% compared with a leading floorplanning method equipped with the proposed pin assignment flow.
\begin{acks}
This work was partially supported by the National Natural Science Foundation of China under Grant No. 62372109 and the Fujian Natural Science Funds under Grant No. 2023J06017.
\end{acks}

\bibliographystyle{unsrt}
\bibliography{sample-base}
\end{document}